\documentclass[prd
 ,twocolumn,a4paper
 ,secnumarabic%
,amssymb,superscriptaddress, amsmath,nobibnotes, aps,showpacs]{revtex4}
\usepackage[T1]{fontenc}
\expandafter\ifx\csname package@font\endcsname\relax\else
 \expandafter\expandafter
 \expandafter\usepackage
 \expandafter\expandafter
 \expandafter{\csname package@font\endcsname}%
\fi

\def\Journal#1#2#3#4{\emph{#1} {\bf #2}, #3 (#4)}

\newtheorem{theorem}{Theorem}[section]
\newenvironment{Theorem}{\begin{theorem}\begin{em}}{\end{em}
\end{theorem}}
\newtheorem{lemma}[theorem]{Lemma}

\newtheorem{corollary}[theorem]{Corrolary}

\newtheorem{conjecture}[theorem]{Conjecture}

\newtheorem{definition}[theorem]{Definition}

\newtheorem{remark}[theorem]{Remark}

\newtheorem{example}[theorem]{Example}

\newtheorem{property}[theorem]{Property}

\newtheorem{proposition}[theorem]{Proposition}
\newenvironment{Proposition}{\begin{proposition}\begin{em}}{\end{em}\end{proposition}}

\def\beq{\begin{eqnarray}}
\def\eeq{\end{eqnarray}}
\def\best{\begin{eqnarray*}}
\def\eest{\end{eqnarray*}}
\def\bcom{}

\def\ee{{\mathbf e}}

\def\upar{\dot{u}_\parallel}

\def\ee{{\mathbf e}}

\def\lc{\overline{\lambda}}
\def\muc{\overline{\mu}}

\def\pic{\overline{\pi}}
\def\kc{\overline{\kappa}}

\def\nuc{\overline{\nu}}
\def\rhoc{\overline{\rho}}
\def\sic{\overline{\sigma}}
\def\sc{\overline{\sigma}}
\def\tauc{\overline{\tau}}
\def\s5{\sqrt{5}}

\def\ah{{\hat{a}}}
\def\bh{{\hat{b}}}
\def\ch{{\hat{c}}}
\def\ddh{{\hat{d}}}
\def\eh{{\hat{e}}}
\def\fh{{\hat{f}}}

\def\di{\textrm{d}}

\def\we{\textrm{w}}

\def\u{{\mathbf u}}
\def\v{{\mathbf v}}
\def\mm{{\mathbf m}}
\def\mmc{\overline{{\mathbf m}}}
\def\kk{{\mathbf k}}
\def\ll{{\mathbf l}}

\def\P{\Psi}

\def\Pc{\overline{\Psi}}

\def\I{\textrm{i}}

\def\a{\alpha}
\def\b{\beta}
\def\g{\gamma}

\def\d{\delta}

\def\t{\theta}

\def\k{\kappa}
\def\l{\lambda}
\def\L{\Lambda}

\def\s{\sigma}

\def\bu{b}
\def\bv{\tilde{b}}

\def\w{\omega}

\def\Zc{\bar{Z}}

\def\mc{\overline{m}}

\def\bar{\overline}

\def\Im{\mbox{Im}}
\def\Re{\mbox{Re}}

\def\tho{\textrm{\TH}}

\begin{document}

\title{Expanding perfect fluid generalizations of the C-metric}

\author{Lode Wylleman\footnote{
Supported by a BOF Research Found (UGent) E-mail: {\tt lode.wylleman@ugent.be}}}
\affiliation{Faculty of Applied Sciences TW16, Ghent University,
Galglaan 2, 9000 Gent, Belgium}
\author{David Beke \footnote{Ph.D. Fellow of the Research Foundation - Flanders (FWO), E-mail: {\tt David.Beke@UGent.be}}}
\affiliation{Faculty of Applied Sciences TW16, Ghent University,
Galglaan 2, 9000 Gent, Belgium}
\affiliation{Centre de Physique Th\'eorique, Campus de Luminy, 13288 Marseille, France}
\date{\today}
\begin{abstract}
Petrov type $D$ gravitational fields, generated by a
perfect fluid with spatially homogeneous energy density and with
flow lines which form a non-shearing and non-rotating timelike
congruence, are re-examined. It turns out that the anisotropic such spacetimes, which
comprise the vacuum C-metric as a limit case, can have
\emph{non-zero} expansion, contrary to the conclusion in the
original investigation by Barnes~\cite{Barnes1}. Apart from the static members, this class consists
of cosmological models with precisely one symmetry. The general line element is constructed and some important
properties are discussed.
It is also shown that purely electric Petrov type $D$ vacuum spacetimes admit
shearfree normal timelike congruences everywhere, even in the non-static regions.
This result incited to deduce intrinsic, easily testable
criteria regarding shearfree normality and staticity of Petrov type
$D$ spacetimes in general, which are added in an appendix.
\end{abstract}
\maketitle

\section{Introduction}\label{section: Introduction}

The C-metric is a well-known exact solution of Einstein's vacuum
equation with zero cosmological constant. The static region of the
corresponding spacetime was first described by Weyl~\cite{Weyl}. At
about the same time Levi-Civita~\cite{LeviCivita} constructed its
line element in closed form, arriving at essentially one cubic
polynomial with two parameters as the metric structure function. The
C-metric is a Petrov type $D$ solution for which at each spacetime
point both Weyl principal null directions (PNDs) are geodesic,
non-shearing, non-twisting but diverging; it thus belongs to the
Robinson-Trautman class of solutions and was rediscovered as
such~\cite{RobTraut}. The label `C' derives from the invariant
classification of static degenerate Petrov type $D$ vacuum
spacetimes by Ehlers and Kundt~\cite{EhlersKundt}. The importance of
this solution as summarized by Kinnersley and
Walker~\cite{KinnersleyWalker}, is threefold.
First, the C-metric describes a spacetime with only two independent Killing
vector fields (KVFs) which can be fully analyzed. Next, it is an `example of
almost everything', most notably it describes a radiative, locally
asymptotically flat spacetime, whilst containing a static region.
The C-metric is contained in the class of boost-rotation-symmetric
spacetimes~\cite{Bicak,Pravda}, which are the only axially
symmetric, radiative and asymptotically flat spacetimes with two
Killing vectors. Finally, the solution has a clear physical
interpretation as the anisotropic gravitational field of two
Schwarzschild black holes
being uniformly accelerated in opposite directions by a cosmic
string or strut, provided that $m\a<1/\sqrt{27}$, where the mass $m$
and acceleration $\alpha$ are equivalents of the two essential
parameters of Levi-Civita~\cite{KinnersleyWalker,Bonnor} (see,
however, the end of $\S$~\ref{subsection:Einstein spaces} for a comment). 

Generalizations of the C-metric have been widely considered. Adding
a cosmological constant $\L$ is straightforward, and we will
henceforth refer with `C-metric' to such Einstein spaces.
Incorporating electromagnetic charge $|q|^2\equiv e^2+g^2$ is equally
natural and leads to quartic structure
functions~\cite{KinnersleyWalker}. Recently, the question how to
include rotation for the holes received a new
answer~\cite{HongTeo2,GriffithsPodolsky}, avoiding the NUT-like
behavior of the previously considered `spinning
C-metric'~\cite{spinCmetric1,spinCmetric2}. All these
generalizations fit in the well-established class ${\cal D}$ of
Petrov type $D$ Einstein-Maxwell solutions with a non-null
electromagnetic field possessing geodesic and non-shearing null
directions aligned with the PNDs~\cite{Debeveretal2,Garcia}, which
reduces for zero electromagnetic field to the subclass ${\cal D}_0$
of Petrov type $D$ Einstein spaces and which contains all well-known
4D black hole metrics. In fact, all ${\cal D}$-metrics can be
derived by performing `limiting contractions'~\cite{SKMHH} from the
most general member, the Plebianski-Demianski line
element~\cite{PlebDem}, which exhibits two quartic structure
functions with six essential parameters $m$, $\alpha$, $|q|^2$, $\L$,
NUT parameter $l$~\cite{NUT} and angular momentum $a$. A physically
comprehensive and simplified treatment can be found in
\cite{Griffiths2}, also surveying recent work in this direction.

In this paper we present a new family of Petrov type $D$, expanding
and anisotropic perfect fluid (PF) generalizations of the C-metric.
The direct motivation and background for this work is the following.

According to the Goldberg-Sachs theorem~\cite{GoldbergSachs} the two
PNDs of any member of ${\cal D}_0$ are precisely those null
directions which are geodesic and non-shearing. Such a member is
purely electric (PE, cf.\ appendix B) precisely
when both PNDs, as well as the complex null directions orthogonal to them, are non-twisting
(non-rotating or hypersurface-orthogonal (HO)).
This is in particular the case for
the C-metric. As we will show, it implies the existence of an {\em
umbilical synchronization} (US), i.e., a non-shearing and
non-rotating unit {\em timelike} vector field (tangent to a
congruence of observers). The importance of USs in cosmology was
stressed in~\cite{Ferrando1992}. If a congruence of observers
measuring isotropic radiation admits orthogonal hypersurfaces, an US
exists. Only small deviations from isotropy are seen in the cosmic
microwave background, and scalar perturbations of a
Friedmann-Lema\^{i}tre-Robertson-Walker universe preserve the
existence of an US~\cite{Ferrando1994}. In general, spacetimes
admitting an US have zero magnetic part of the Weyl tensor wrt
it~\cite{Trumper} and thus
are either of Petrov type $O$, or PE and of type
$D$ or $I$~\cite{SKMHH}. Conformally flat spacetimes always admit
USs (see e.g.\ (6.15) in \cite{SKMHH}). Tr\"{u}mper showed that
algebraically general vacua with an US are static~\cite{Trumper1}.
Motivated by this result and by his own work \cite{Barnes2} on
static PFs, Barnes~\cite{Barnes1} studied PF spacetimes with an US
tangent to their flow lines. He was able to generalize Tr\"{u}mper's
result to Petrov type $I$ such PFs and recovered Stephani's results on conformally
flat PF solutions which are either of generalized Schwarzschild type
or of generalized Friedmann type (so called Stephani
universes)~\cite{Stephani1}.
The type $D$ solutions were integrated and invariantly
partitioned, based on the direction of the gradient of the energy
density relative to the PNDs and the flow vector at each point.
Class I, characterized by the energy density being constant on the hypersurfaces orthogonal to the
flow lines and thus the only class containing Einstein spaces as
limit cases, was further subdivided using the gradient of $\Psi_2$
(cf.\ $\S$~\ref{subsection:classification} for details). By
solving the field equations, Barnes concludes that class ID,
consisting of the anisotropic class I models, solely contains
non-expanding solutions. Hence, these PF solutions would not be
viable as a cosmological model. However, based on an integrability
analysis of class I in the Geroch-Held-Penrose (GHP)
formalism~\cite{GHP}, we found that this conclusion cannot be valid
and this led to a detailed reinvestigation.

In this article we construct the general line element of the full ID class, comprising both the known
non-expanding perfect fluid models and the new expanding ones, and discuss some
elementary properties. We want to stress the following point. The full class represents a PF
generalization of the C-metric in the sense that the C-metric is
contained as the Einstein space limit. The physical interpretation
of this fact is however not established. This would require to
exhibit this solution for small masses as a perturbation of a known
PF solution, just as the C-metric interpretation of small
accelerating black holes has been established in a flat or (anti-)de
Sitter
background~\cite{KinnersleyWalker,Podol1,Podol2,Dias,Emparan}.

However, the mathematical relation with the C-metric is useful. As
already deduced in \cite{Barnes1}, the PF solution is, just as the
C-metric, conformally related to the direct sum of two 2D metrics.
The fact that one part is equal for the PF solution and the C-metric
is helpful in the analysis, e.g. we will show that (a part of) the
axis of symmetry can readily be identified as a conical singularity,
analogous to the defect of the cosmic string present in the C-metric. The
non-static spacetimes presented are exact perfect fluid solutions
with only this symmetry, and the analysis appears to be within
reach. For the expanding ID PF models both the matter density $w(t)$
and the expansion scalar $\theta(t)$ can be arbitrary functions.
This freedom is displayed explicitly in the metric form, and makes
the solutions more attractive as a cosmological model.

The paper is organized as follows. In section 2 we present the GHP
approach to class I. We derive a closed set of equations, construct
suitable scalar invariants, interpret the invariant
subclassification of \cite{Barnes1} and start the
integration. At the end we provide alternative characterizations for
the Einstein space members and identify their static regions and USs. In section 3 we finish the
construction of the general ID line element in a
transparent way, and correct the calculative error of
\cite{Barnes1} in the original approach. Then we deduce basic
properties of the ID perfect fluid models. In section 4 we summarize the main results
and indicate points of further research.
The work greatly benefited from the
use of the GHP formalism, which at the same time elucidates the
deviation from the C-metric. In appendix A we provide a pragmatic survey of this formalism for the non-expert reader.
In appendix B, finally, we present criteria for deciding when a Petrov
type $D$ spacetime admits a (rigid) US or is static.

{\em Notation.} For spacetimes $(M,g_{ab})$ we take (+ + + -) as the
metric signature and use geometrized units $8\pi G=c=1$, where $G$
is the gravitational coupling constant and $c$ the speed of light.
$\Lambda$ denotes the cosmological constant. We make consistent use
of the abstract Latin index notation for tensor fields, as advocated
in \cite{Penrose1}. Round (square) brackets denote
(anti-)symmetrization, $\eta_{abcd}$ is the spacetime alternating
pseudo-tensor  and $ \nabla_c T_{ab\ldots}$ (${\cal L}_{\bf X} T_{ab\ldots}$) designates the
Levi-Civita covariant derivative (Lie derivative wrt $X^a$) of the tensor field $T_{ab\ldots}$.
One has
\begin{equation*}
\di_a f=\nabla_af,\quad \di_b Y_a=\nabla_{[b} Y_{a]}
\end{equation*}
for the exterior derivative of a scalar field $f$, resp.\ one-form
field $Y_a$, and
\begin{equation*}
\mathbf{X}(f)
\equiv X^a\di_a f
\end{equation*}
denotes the Leibniz action of a vector field $X^a$ on $f$; when $X^a$ is the
$x^i$-coordinate vector field $\partial_{x^i}{}^a$ we write $\partial_{x^i}f$ or $f_{,{x^i}}$, and a prime denotes ordinary derivation for functions of one variable, $f'(x)\equiv \partial_x f(x)$. However, we use
index-free notation in line elements $\di s^2=g_{ij}\di x^i\di x^j$.
The specific GHP notation is introduced in appendix A.

\section{GHP approach to class I}\label{section: GHP approach}

\subsection{Definition and integrability}\label{subsection:Integrability}

We consider Barnes' class I~\cite{Barnes1}, consisting of spacetimes
$(M,g_{ab})$ with the following properties:
\begin{enumerate}
\item[(i)] the spacetime admits a unit timelike
vector field $u^a$ ($u^au_a=-1$) which is non-shearing and
non-rotating, i.e., its covariant derivative is of the form
\begin{equation}\label{ua;b}
\nabla_b u_{a}=\theta h_{ab}-\dot{u}_au_b,\quad h_{ab}\equiv
g_{ab}+u_au_b,
\end{equation}
where the acceleration $\dot{u}^a=u^b\nabla_b u^a$ and expansion
rate $\theta=\nabla_a u^a$ are the remaining kinematic quantities of
$u^a$;
\item[(ii)] the Einstein tensor has the structure
\begin{eqnarray}
&&G_{ab}=Su_au_b+pg_{ab}=wu_au_b+ph_{ab},\\
&&D_a w\equiv h_a{}^b\nabla_b w=0,
\end{eqnarray}
i.e., the spacetime represents the gravitational field of either a
perfect fluid with shearfree normal
four-velocity $u^a$, pressure $p+\Lambda$ and spatially homogeneous
energy density $w-\Lambda$ (case $S\equiv w+p\neq 0$) or a vacuum
(\emph{Einstein space} case $S=0$, where $w=-p$ may be identified
with $\Lambda$);
\item[(iii)] the Weyl tensor $C_{abcd}$ is degenerate but non-zero, i.e., the spacetime is algebraically special but not conformally flat.
\end{enumerate}
Choose null vector fields $k^a$ and $l^a$, subject to the normalization condition $k^al_a=-1$, such that
\begin{eqnarray}
u^a=\frac{1}{\sqrt{2q}}\left(qk^a+l^a\right),\quad q>0.\label{udef}
\end{eqnarray}
Within the GHP formalism (cf.\ appendix A) based on the complex null tetrad $(k^a,l^a,m^a,\mc^a)$, $q$ is (-2,-2)-weighted
and the conditions (i) and (ii) translate into
\begin{eqnarray}
&&\pi+\tauc=q\kc+q^{-1}\nu,\;\;\;\l=q\sic,\;\;\;\mu-\muc=q(\rhoc-\rho),\;\;\;\;\;\label{lambda/sigma pi/tau}\\
&&\tho'q-q\tho q=-2q(\mu-q\rhoc),\quad\eth q=\eth'q=0\label{deriv q}
\end{eqnarray}
and
\begin{eqnarray}
&&\Phi_{01}=\Phi_{12}=\Phi_{02}=0,\label{Phi_ij 0}\\
&&\Phi_{11}=\frac{S}{8},\quad \Phi_{00}=\frac{S}{4q},\quad \Phi_{22}=\frac{qS}{4},\label{Phi_ii}\\
&&R\equiv 24\Pi=w-3p=4w-3S,\label{Ricciscalar}\\
&& \eth w=\;\eth' w\,=0,\quad \tho'w-q\tho w=0,\label{deriv w}
\end{eqnarray}
respectively. By virtue of condition (i) the magnetic part $H_{ab}\equiv\frac{1}{2}\eta_{acmn}C^{mn}{}_{bd}u^cu^d$ of the Weyl tensor wrt $u^a$ vanishes~\cite{Trumper}. 
In combination with condition (iii) it follows that the Weyl tensor is purely electric wrt $u^a$, $E_{ab}\equiv C_{acbd}u^cu^d\neq 0$, the Weyl-Petrov type is $D$,
and at each point $u^a$ lies in the plane $\Sigma$ spanned by the Weyl
PNDs (cf.\ appendix B for a GHP proof of these well-known facts).
Hence, choosing $k^a$ and $l^a$ along the PNDs, $(k^a,l^a,m^a,\mc^a)$ is a Weyl principal null tetrad (WPNT) and we have
\begin{eqnarray}
&&\Psi_0=\Psi_1=\Psi_3=\Psi_4=0,\label{Psis}\\
&&\Pc=\Psi\neq 0, \quad \Psi\equiv 2\Psi_2.\label{Psi2 real}
\end{eqnarray}
Under the restrictions (\ref{Phi_ij 0}) and (\ref{Psis}), the GHP Bianchi equations are given by (\ref{bi1})-(\ref{bi6}) and their prime duals. Combining these with the other equations in (\ref{lambda/sigma pi/tau})-(\ref{Psi2 real}) results in
\begin{eqnarray}
&&\k=0,\quad \nu=0,\quad \s=\l=0,\label{geod shearfree}\\
&&\rhoc=\rho,\quad\muc=\mu,\quad\pi=-\tauc,\label{pitau}\\
&&\tho\P=3\rho\P,\quad \tho'\P=-3\mu\P,\label{thornPsi}\\
&&\eth \P=3\tau\P,\quad \eth'\P=-3\pi\P,\label{ethPsi}\\
&&\tho'S-q\tho S=S(\tho q-\mu+q\rho),\label{thoS}\\
&&\eth S=\tau S,\quad\eth'S=\tauc S,\label{ethS}\\
&&\tho' w=q\tho w=-\frac{3S(\mu-q\rho)}{2}.\label{thow}
\end{eqnarray}
With (\ref{Phi_ij 0})-(\ref{Ricciscalar}) and (\ref{Psis})-(\ref{pitau}) the Ricci equations, given by (\ref{ghp1})-(\ref{ghp6}) and their prime duals, reduce to
\begin{eqnarray}
&&\tho\mu=-\tho'\rho\label{thomu}\\
&&\quad\,\,\,=-\eth'\tau+\mu\rhoc+\tau\tauc+\frac{\P}{2}+\frac{w}{3}-\frac{S}{4},\label{tho'rho}\\
\label{derivmu}&&\tho'\mu=-\mu^2-\frac{qS}{4},\quad
\eth\mu=\eth'\mu=0,\\
\label{derivrho}&&\tho\rho=\rho^2+\frac{S}{4q},\quad
\eth\rho=\eth'\rho=0,\\
&&\tho\tau=\tho'\tau=0,\quad \eth\tau=\tau^2,\label{derivtau}\\
&&-\eth\pi=\eth'\tau=\eth\tauc\equiv \frac{H}{2}\label{def Z1}
\end{eqnarray}
and the complex conjugates of (\ref{derivtau}),
while the commutator relations applied to a $(\we_p,\we_q)$-weighted
scalar $\eta$ become
\begin{eqnarray}
&&{}[\tho,\tho']\eta=(\we_p+\we_q)\left(\tau\tauc-\frac{\Psi}{2}+\frac{w}{6}-\frac{S}{4}\right)\eta,\label{d34}\\
&&{}[\eth,\eth']\eta=(\we_p-\we_q)\left(-\mu\rho+\frac{\Psi}{2}-\frac{w}{6}\right)\eta,\label{d12}\\
&&{}[\tho,\eth]\eta=(-\tau\tho+\rho\eth+\we_q\,\rho\tau)\eta,\label{d14}\\
&&{}[\tho,\eth']\eta=(-\tauc\tho+\rho\eth'+\we_p\,\rho\tauc)\eta,\label{d24}\\
&&{}[\tho',\eth]\eta=(-\tau\tho'-\mu\eth+\we_p\,\mu\tau)\eta,\label{d13}\\
&&{}[\tho',\eth']\eta=(-\tauc\tho'-\mu\eth'+\we_q\,\mu\tauc)\eta.\label{d23}
\end{eqnarray}
Then the $[\eth,\eth'](\tau)$, $[\eth,\eth'](\tauc)$,
$[\tho,\eth'](\tau)$ and $[\tho',\eth'](\tau)$ commutator relations
imply
\begin{eqnarray}
&&\eth H=2\tau(H+\P-G),\quad \eth'H=2\tauc(H+\P-G),\nonumber\\
&&\tho H=\rho(H+F),\quad \tho'H=-\mu(H+F),
\label{derivethtau}
\end{eqnarray}
where
\begin{equation}
F\equiv 2\tau\tauc,\quad G\equiv 2\mu\rho+\frac{w}{3}.\label{def FG}
\end{equation}
One checks that the integrability conditions for the system
(\ref{deriv q})-(\ref{def FG}) of partial differential equations (PDEs)
are identically satisfied,
indicating that corresponding solutions exist. Those for which $u^a$
is non-expanding additionally satisfy
\begin{equation}\label{mu=qrhoc}
\theta\sim \mu-q\rhoc=0
\end{equation}
(cf.\ (\ref{def bu}) and (\ref{def bu bv}) below). However, (\ref{mu=qrhoc}) does not
follow as a consequence of the ans\"{a}tze; this implies the
existence of expanding anisotropic perfect fluid models in class I
($\S$ \ref{section:pfmetric}). Also, the scalar invariant $\mu\rho$ may be strictly negative, which is incompatible with
(\ref{mu=qrhoc}); as a consequence, the class I Einstein spaces are not necessarily static ($\S$~\ref{subsection:Einstein spaces}).

\subsection{Metric structure and
subclassification}\label{subsection:classification}

The first, second and last parts of (\ref{geod
shearfree})-(\ref{pitau}) precisely account for the
hypersurface-orthogonality of $k^a$, $l^a$ and $m^a\leftrightarrow \mc^a$, respectively.
Thus real scalar fields $u$, $v$, (zero-weighted) and $U$, $V$
($(-1,-1)$- resp.\ $(1,1)$-weighted), and complex scalar fields
$\zeta$ (zero-weighted) and $Z$ ((1,-1)-weighted) exist such that
\begin{eqnarray}
\di_a u=\frac{\P^{1/3}}{U}k_a,\quad \di_a
v=\frac{\P^{1/3}}{V}l_a,\quad \di_a \zeta=\frac{\P^{1/3}}{Z}m_a.
\label{dzeta}
\end{eqnarray}
By (\ref{dfweighted}) this is equivalent to
\begin{eqnarray}
\tho' u=-\P^{1/3}/U,\quad \tho u=\eth u=\eth'u=0,\label{deriv a}\\
\tho v=-\P^{1/3}/V,\quad \tho' v=\eth v=\eth'v=0,\\
\eth'\zeta=\P^{1/3}/Z,\quad \tho\zeta=\tho'\zeta=\eth\zeta=0,\\
\eth\bar{\zeta}=\P^{1/3}/\bar{Z},\quad
\tho\bar{\zeta}=\tho'\bar{\zeta}=\eth'\bar{\zeta}=0.\label{deriv
zetac}
\end{eqnarray}
The commutator relations (\ref{d14})-(\ref{d23}) applied to $u$,
$v$, $\zeta$ and $\bar{\zeta}$ then yield
\begin{eqnarray}
&&\eth U=\eth'U=\eth V=\eth'V=0,\label{deriv AB}\\
&& \tho Z=\tho'Z=\tho \bar{Z}=\tho'\bar{Z}=0.\label{deriv P}
\end{eqnarray}
Hence, when we take these fields as coordinates,
(\ref{dzeta})-(\ref{deriv P}) imply that the zero-weighted fields
$UV$ and $Z\bar{Z}$ only depend on $(u,v)$, resp.\
$(\zeta,\bar{\zeta})$, such that all class I metrics are conformally
related to direct sums of metrics on two-spaces:
\begin{eqnarray}
g_{ab}&=&\Psi^{-2/3}(g^\perp_{ab}\oplus g^\Sigma_{ab}),\label{ds2HO}\\
g^\perp_{ab}&\equiv&
2\Psi^{2/3}m_{(a}\mc_{b)}=2Z\bar{Z}(\zeta,\bar{\zeta})\,\di_{(a}\zeta\,\di_{b)}\bar{\zeta},\label{ds2perp}\\
g^\Sigma_{ab}&\equiv &
-2\Psi^{2/3}k_{(a}l_{b)}=-2UV(u,v)\,\di_{(a}u\,\di_{b)}v.\label{ds2sigmakl}
\end{eqnarray}
The line elements of $g^\perp_{ab}$ and $g^\Sigma_{ab}$ will be
denoted by $\di s_{\perp}^2$, resp.\ $\di s_{\Sigma}^2$.

In the case where such a two-space is not of constant curvature,
however, we will construct more suitable coordinates in the sequel.
Inspired by the GHP manipulations of \cite{Edgar type D}
for type $D$ vacua~\footnote{More precisely equations (48)-(53) of ref.\ \cite{Edgar type D}. There are some typos in these equations:
the integers 3 and 9 should be omitted in (50) and (51)-(52),
respectively, while there should be $9k$ instead of $k$ in (53).},
we start this construction by deducing suitable combinations of the
scalar invariants $F$, $G$, $H$ and $\Psi$. From (\ref{dfweighted}),
(\ref{deriv w}) and (\ref{thornPsi})-(\ref{def FG}) it
is found that
\begin{eqnarray}
&\di_a F=3\P^{1/3}\varphi\,\a_a,&\quad \di_a G=3\P^{1/3}\g\,\b_a,\label{dFG}\\
&\di_a \varphi=2\P^{1/3}x\,\a_a,&\quad \di_a \g=2\P^{1/3}y\,\b_a,\label{drs}\\
&\di_a x=\P^{1/3}\,\a_a,&\quad \di_a y=\P^{1/3}\,\b_a,\label{dxy}
\end{eqnarray}
where
\begin{eqnarray}\label{def alpha_a beta_a}
\a_a\equiv \tauc m_a+\tau\mc_a,\quad \beta_a\equiv \mu k_a-\rho l_a
\end{eqnarray}
are invariantly-defined one-forms and
\begin{eqnarray}
\varphi\equiv \frac{H+F}{3\P^{1/3}},&& \g\equiv
\frac{-H+\P+F+2G}{3\P^{1/3}},\label{def uv}\\
x\equiv \frac{H+\P-G}{3\P^{2/3}},&& y\equiv
\frac{-H+2\P+G}{3\P^{2/3}}.\label{def xy}
\end{eqnarray}
Consequently, the scalar invariants
\begin{eqnarray}
&&C\equiv 3(\varphi-x^2)=3(\gamma-y^2),\label{def C}\\
&&D\equiv -x^3-Cx+F=y^3+Cy-G.\label{def D}
\end{eqnarray}
are {\em constant} ($\di_a C=\di_a D=0$). From (\ref{def xy}) and
(\ref{def D}) it follows that $F$, $G$, $H$ and $\Psi$ are
biunivocally related to
$x$, $y$,
$C$ and $D$, where
\begin{eqnarray}
&&2\tau\tauc\equiv F=x^3+Cx+D,\label{def F}\\
&&2\mu\rho\equiv G-\frac{w}{3}=y^3+Cy-D-\frac{w}{3},\label{def G}\\
&&2\eth'\tau\equiv H=2x^3+3x^2y+Cy-D,\label{def H}\\
&&\Psi=(x+y)^3\neq 0.\label{def Psi}
\end{eqnarray}

Barnes~\cite{Barnes1} partitioned class I according to the position
of the gradient $\nabla^a\Psi$ relative to $\Sigma$ and
$\Sigma^\perp$. This relates to the vanishing of the invariants
$\tau\tauc=-\pi\tau$ or $\mu\rho$, maximal symmetry of
$g^\perp_{ab}$ or $g^\Sigma_{ab}$ and spatial rotation or boost
isotropy of $g_{ab}$, as follows.

First assume $\tau=0$. In this case (\ref{def Z1})
and the first parts of (\ref{def FG}) and (\ref{def uv})-(\ref{def
D}) imply
\begin{eqnarray}
&&H=F=\varphi=0,\quad \Psi-G=3x\Psi^{2/3},\nonumber\\
&&C=-3x^2,\quad D=2x^3,\label{tau=0 conseq}
\end{eqnarray}
such that $x$ is constant. In combination with the last part of (\ref{pitau}), (\ref{ethPsi}) and the first
parts of (\ref{dxy})-(\ref{def alpha_a beta_a}) one gets
\begin{eqnarray}
\tau\tauc=0 \Leftrightarrow \pi=\tau=0 \Leftrightarrow x\,=\,const
\Leftrightarrow \nabla^a\Psi\in\Sigma.\label{LRS karakt}
\end{eqnarray}
The $[\eth,\eth']$ commutator relation applied to $\zeta$,
$\bar{\zeta}$ and $Z$ imply $\eth Z=\eth'\Zc=0$ and
$\eth\eth'Z=3x\Psi^{2/3}Z$. Herewith the Gaussian curvature of the
two-space with metric $g^\perp_{ab}$ becomes
\begin{eqnarray}
K^\perp&=&-(Z\Zc)^{-1}(\ln(Z\Zc))_{,\zeta\bar{\zeta}}
=-\Psi^{-2/3}\eth\eth'(\ln Z\Zc)\nonumber\\
&=&-\Psi^{-2/3}\eth\left(\frac{\eth' Z}{Z}\right) =-3x,\label{Kperp}
\end{eqnarray}
where the dual of (\ref{dzeta}) was used in the calculation. In
conjunction with the results of Goode and Wainwright~\cite{Goode},
we conclude that (\ref{LRS karakt}) yields the class I solutions
which are locally rotationally symmetric (LRS) of label II in the
Stewart-Ellis classification~\cite{StewartEllis}, characterized by
$g^\perp_{ab}$ having constant curvature $K^\perp=-3x$. As well
known (see e.g.\ the appendix of \cite{Szekeres}) the coordinates
$\zeta$ and $\bar{\zeta}$ may then be adapted such that
$Z\bar{Z}(\zeta,\bar{\zeta})=(1+K^\perp\zeta\bar{\zeta}/2)^{-1}$ in
(\ref{ds2perp}), or an alternative form may be taken:
\begin{eqnarray}
&&\di s_\perp^2=\frac{2\di \zeta \di
\bar{\zeta}}{1+\frac{K^\perp}{2}\zeta\bar{\zeta}}=Y_\bot^2(\di
x_1^2+\cos(\sqrt{k_\bot}x_1)^2\di x_2^2),\nonumber\\
&&\label{ds2perp constcurv}K^\bot=k_\bot Y_\bot^{-2},\quad
k_\perp\in\{-1,0,1\}.
\end{eqnarray}

Now assume $\mu\rho=0$. It follows from
(\ref{thomu})-(\ref{derivrho}), (\ref{def FG}), (\ref{def H}) and the second parts of  (\ref{def uv})-(\ref{def
D}) that
\begin{eqnarray}
&&S=\gamma=0,\quad G=\frac{w}{3}\equiv\frac{\L}{3},\quad -H+2\Psi+\frac{\L}{3}=3y\Psi^{2/3},\nonumber\\
&& C=-3y^2,\quad
D=-2y^3-\frac{\L}{3},\label{mu=0conseq}
\end{eqnarray}
such that $y$ is constant.
In combination with (\ref{thornPsi}) and the second parts of
(\ref{dxy})-(\ref{def alpha_a beta_a}) this implies
\begin{eqnarray}
\mu\rho=0 \Leftrightarrow \mu=\rho=0 \Leftrightarrow y\,=\,const
\Leftrightarrow \nabla^a\Psi\in\Sigma^\perp.\label{boost karakt}
\end{eqnarray}
By a similar reasoning as in the case $\tau=0$ one concludes that
(\ref{boost karakt}) yields the locally boost-isotropic Einstein
spaces of Petrov type $D$, characterized by $g^\Sigma_{ab}$ having
constant curvature
\begin{equation}\label{KSigma}
K^\Sigma=-3y,
\end{equation}
such that in this case one may
take $UV(u,v)=(1-K^\Sigma uv/2)^{-1}$ in (\ref{ds2sigmakl}) and we
have
\begin{eqnarray}
&&\di s_\Sigma^2=-\frac{2\di u \di
v}{1-\frac{K^\Sigma}{2}uv}=Y_\Sigma^2(\di
x_3^2-\cos(\sqrt{k_\Sigma}x_3)^2\di x_4^2),\nonumber\\
&&K^\Sigma=k_\Sigma Y_\Sigma^{-2},\quad
k_\Sigma\in\{-1,0,1\}.\label{ds2Sigma constcurv}
\end{eqnarray}
With (\ref{ds2HO}) and $\di s_\Sigma^2$ written in the second form, it is clear
that
\begin{equation}\label{KVF x4}
\partial_{x_4}{}^a=-\Psi^{-2/3}Y_\Sigma^2\cos(\sqrt{k_\Sigma}x_3)^2\di^a x_4
\end{equation}
is a HO timelike Killing vector field.

Four subclasses of class I thus arise, which were labeled by Barnes
as follows:
\begin{eqnarray}\label{Barnes division}
\begin{array}{cc}
\mbox{IA}:\,\, \tau=0=\mu\rho,\quad& \mbox{IB}:\,\,
\tau=0\neq \mu\rho,\\
\mbox{IC}:\,\, \tau\neq 0=\mu\rho,\quad& \mbox{ID}:\,\, \tau\neq
0\neq \mu\rho.
\end{array}
\end{eqnarray}
We proceed with the respective integrations. Notice that in the
joint case $\mu\rho\tau=0$ one has
\begin{eqnarray}
2(\tau\tauc+\mu\rho)=(x+y)^3+K(x+y)^2-\frac{w}{3},\label{taumurhospec}
\end{eqnarray}
with $K=K^\perp$ for $\tau=0$ and $K=K^\Sigma$ for $\mu\rho=0$. When
$\tau\neq 0$ or $\mu\rho\neq 0$ we may take $x$, resp.\ y as a
coordinate, where (\ref{dxy})-(\ref{def alpha_a beta_a}) and
(\ref{def Psi}) imply
\begin{eqnarray}
(x+y)(\tauc m_a+\tau \mc_a)=\di_a x,\label{invform dx}\\
(x+y)(\mu k_a-\rho l_a)=\di_a y.\label{invform dy}
\end{eqnarray}
In view of (\ref{ds2HO})-(\ref{ds2sigmakl}) and (\ref{def Psi}) it
then remains to determine suitable complementary coordinates for $x$ in
$g_{ab}^\perp$ or $y$ in $g_{ab}^\Sigma$.

For $\tau\neq 0$, Frobenius' theorem and (\ref{invform dx}) suggest
to examine whether zero-weighted functions $\phi$ and $f$ exist such
that
\begin{equation}\label{dphi}
\I\frac{x+y}{2\tau\tauc}\left(\tau\mc_a-\tauc m_a\right)=f\di_a
\phi.
\end{equation}
By (\ref{dfweighted}) this amounts to calculating the integrability conditions of the
system
\begin{equation}\label{Phi system}
\tho \phi=\tho'\phi=0,\quad
\tauc\eth\phi=-\tau\eth'\phi=\I\frac{x+y}{2f}
\end{equation}
which turn out to be
\begin{equation}\label{df}
\tho f=\tho' f=0,\quad 
\alpha_a\nabla^a f =0.
\end{equation}
These last equations have the trivial solution $f=1$, for which a
solution $\phi$ of (\ref{Phi system}) is determined up to an
irrelevant constant. Herewith the invariantly-defined one-form on
the left-hand side in (\ref{dphi}) is exact,
and we take $\phi$ as the coordinate complementary
to $x$. On solving (\ref{invform dx}) and (\ref{dphi}) with $f=1$
for $m_a$ and $\mc_a$ and using (\ref{def F}) we conclude from (\ref{ds2perp}) that
\begin{equation}
\di s_\perp^2=\frac{\di x^2}{2\tau\tauc}+2\tau\tauc\di \phi^2,\quad
2\tau\tauc=x^3+Cx+D\label{ds2perp nonconst}
\end{equation}
for classes IC and ID. Clearly, the metric solutions should be restricted to spacetime
regions where $x^3+Cx +D>0$ for consistency,
while
\begin{equation}\label{KVF phi}
\partial_\phi{}^a=\I\frac{\tau\mc^a-\tauc
m^a}{x+y}=\frac{2\tau\tauc}{(x+y)^2}d^a\phi,
\end{equation}
is a HO spacelike Killing vector field (KVF).

For $\mu\rho\neq 0$ one analogously considers
\begin{equation}\label{psi system}
\eth\psi=\eth'\psi=0,\quad \mu\tho\psi=\rho\tho'\psi=\frac{x+y}{2g}
\end{equation}
but the integrability conditions of this system are now
\begin{eqnarray}\label{g system}
\eth g=\eth'g=0,\quad \beta_a\nabla^a g
=-gS\frac{\mu^2+q^2\rho^2}{q\mu\rho}.
\end{eqnarray}
So $g=1$ is \emph{only} a solution in the Einstein subcase $S=0$,
for which we then get
\begin{equation}
\di s_\Sigma^2=\frac{\di y^2}{2\mu\rho}-2\mu\rho\di \psi^2,\quad
2\mu\rho=y^3+Cy-D-\frac{\L}{3}\label{ds2Sigma nonconst}
\end{equation}
from (\ref{ds2sigmakl}), (\ref{def G}), (\ref{invform dy}) and (\ref{psi system}), with KVF
\begin{equation}\label{KVF psi}
\partial_\psi{}^a=\frac{\mu k^a+\rho l^a}{x+y}=-\frac{2\mu\rho}{(x+y)^2}\di^a\psi,
\end{equation}
which is timelike for $\mu\rho>0$ and spacelike for $\mu\rho<0$. In
general, the second vector field in (\ref{KVF psi}) is always HO:
the integrability conditions of (\ref{g system}) are checked to be
identically satisfied, such that solutions $g$ and a corresponding
solution $\psi$ of (\ref{psi system}) exist. However, taking $\psi$
as a complementary coordinate of $y$ eventually leads to a very
complicated system of coupled partial differential equations for
$g=g(y,\psi)$, which is impossible to solve explicitly. We shall
remedy this in section \ref{subsection: ID line element} but now
discuss characterizing
features of the Einstein space limit cases.

\subsection{Characterizations of PE Petrov type $D$ Einstein spaces}\label{subsection:Einstein spaces}

Petrov type $D$ Einstein spaces constitute the class ${\cal
D}_0$ (cf.\ the introduction) and are all explicitly known.
The line elements are obtained by putting the electromagnetic
charge parameter $\Phi_0$ or
$e^2+g^2$ equal to zero in the ${\cal D}$-metrics given by Debever
{\em et al.} \cite{Debeveretal2}, resp.\ Garc\'{i}a~\cite{Garcia}. These
coordinate forms generalize and streamline those found by
Kinnersley~\cite{Kinnersley} in the $\L=0$ case.

Recently, a manifestly invariant treatment of ${\cal D}_0$, making
use of the GHP formalism, was presented~\cite{Edgar type D}. Within
GHP, ${\cal D}_0$-metrics are characterized by the existence of a
complex null tetrad wrt which (\ref{Psis}) and $\Phi_{ij}=0$ hold (i.e.,
the tetrad is a WPNT and (\ref{Phi_ij 0})-(\ref{Phi_ii}) with $S=0$
hold). According to the Goldberg-Sachs theorem~\cite{GoldbergSachs}, (\ref{geod
shearfree}) holds and characterizes WPNTs as well. The scalar
invariant identities (see \cite{CzaporMcLen,Edgar type D})
\begin{equation}
\mu\rhoc=\muc\rho,\quad \pi\pic=\tau\tauc,
\end{equation}
just as (\ref{thornPsi})-(\ref{ethPsi}),
(\ref{thomu})-(\ref{tho'rho}) and the first equation of (\ref{def
Z1}) are also valid in general. From these relations it follows that
\begin{equation}\label{Psi2 real equivalence}
(\ref{Psi2 real})\,\, \Leftrightarrow\,\,(\ref{pitau}),
\end{equation}
i.e., a Petrov type $D$ Einstein space is PE if and only if the
WPNT directions are HO. In fact, it can readily be shown by a more
detailed analysis than in \cite{Edgar type D} that if the spacetime
belongs to Kundt's class, i.e., if one of the PNDs is moreover
non-diverging, one has
\begin{eqnarray}\label{mu=0}
\mu=0&\Rightarrow&\rho=0\,\,\mbox{or}\,\,\rho-\rhoc\neq 0\neq
\pi+\tauc.
\end{eqnarray}
Equations (4), (5) and (31) in \cite{Edgar type D} then imply
\begin{eqnarray}
&\rho=\rhoc\neq 0&\Rightarrow\,\,\mu=\muc\neq 0=\pi+\tauc,\label{rho real}\\
&\pi=-\tauc\neq
0&\Rightarrow\,\,\mu=\muc,\,\rho=\rhoc.\label{pi+tauc=0}
\end{eqnarray}
One concludes that the Kundt and Robinson-Trautman subclasses of
${\cal D}_0$ have empty intersection, and that the latter consists of PE spacetimes for which {\em both} PNDs are non-twisting but diverging.
These results - which remain valid for the electrovac class ${\cal D}$, just as the two theorems below -  are implicit in \cite{Debeveretal2}, where
the concerning PE metrics form the Einstein space subclasses of the
classes labeled by 
\begin{eqnarray}\label{Debever division}
\begin{array}{cc}
C^{00}:\,\, \tau=0=\mu\rho,\quad& C^0{}_+:\,\,
\tau=0\neq \mu\rho,\\
C^0{}_-:\,\, \tau\neq 0=\mu\rho,\quad& C^*:\,\, \tau\neq
0\neq \mu\rho.
\end{array}
\end{eqnarray}
By the Einstein space specifications $S=0$ and $w=\L=const$, the boost-field $q$ disappears from the equations (\ref{Phi_ij 0})-(\ref{def FG}) and is not
determined by the geometry, in contrast to the situation for
perfect fluids $S\neq0$ (cf.\ $\S$~\ref{subsection: ID line element}). Moreover, the set
(\ref{lambda/sigma pi/tau})-(\ref{deriv q}), i.e.\ the requirement
that an US given by (\ref{udef}) exists, is decoupled from
(\ref{Phi_ij 0})-(\ref{def FG}) and is not needed to derive
(\ref{thornPsi})-(\ref{def FG}) from (\ref{Phi_ij 0})-(\ref{pitau}).
From the integrability of the complete set (\ref{deriv
q})-(\ref{def FG}), (\ref{Barnes division}) and the above we conclude:
\begin{Theorem}\label{Theorem 1}
The closed set
(\ref{Phi_ij 0})-(\ref{def FG}) characterizes the class ${\cal D}_0$ of
PE Petrov type $D$ Einstein spaces, which are precisely
those Einstein spaces for which the WPNT directions are HO or, alternatively, those which belong
to Barnes' class I, all admitting a
one-degree freedom of USs in all regions of spacetime. Barnes' boost-isotropic Kundt classes IA and IC
coincide with $C^{00}$, resp.\ $C^0{}_-$, while the Robinson-Trautman members of ${\cal D}_0$ constitute $C^0{}_+$ and $C^*$, form
the Einstein space subclasses of IB, resp.\ ID, and possess non-twisting but diverging PNDs at each point.
\end{Theorem}
The result is in agreement with proposition \ref{Prop: US}, which provides criteria for deciding when a Petrov type $D$ spacetime allows for an US,
regardless of the structure of the energy-momentum tensor. The hypersurface-orthogonality
(\ref{geod shearfree})-(\ref{pitau}) of
the WPNT directions corresponds to criterion 5 and is actually equivalent to a one-degree freedom of USs.
It is worth to mention that all LRS II spacetimes, i.e.\ those exhibiting (pseudo-)~spherical or plane
symmetry, share this property with the ${\cal D}_0$- and ${\cal D}$-metrics. On the other hand, certainly not all PE spacetimes admit an US. For instance, the G\"{o}del solution is an LRS I PE perfect fluid of Petrov type $D$, described in GHP by (\ref{Phi_ij 0})-(\ref{geod shearfree}) and
\begin{eqnarray}
&&S/2=w=p=-3\Psi_2=-2\mu\rho=\mbox{const.}>0,\\
&&\pi=\tau=0,\quad \mu=q\rho,\quad \muc=-\mu,\quad\rhoc=-\rho,
\end{eqnarray}
$q>0$ being annihilated by all weighted GHP-derivatives; hence
the invariant $(\mu-\muc)(\rho-\rhoc)=4\mu\rho=4q\rho^2$ appearing in criterion 2 of proposition \ref{Prop: US} is strictly negative, and it follows that the G\"{o}del solution does not admit an US.
As another example, the spatially-homogeneous $\L=0$ vacuum metrics
\begin{eqnarray}\label{Kasnermetric}
&&\di s^2=t^{2p_1}\di x^2+t^{2p_2}\di y^2+t^{2p_3}\di z^2-\di t^2,\\
&&p_1+p_2+p_3=p_1^2+p_2^2+p_3^2=1,\quad p_1p_2p_3\neq 0.
\end{eqnarray}
attributed to Kasner~\cite{Kasner} are PE~\footnote{If one of the $p_i$'s is zero then so is a second one; this reproduces Minkowski.}. If the $p_i$ are all different, there is a complete group $G_3 I$ of isometries and the Petrov type is I. In this case $\partial_t{}^a$ is the up to reflection unique Weyl principal vector field and hence the only possible US-candidate; however, its shear tensor
has the non-zero eigenvalues $(1/3-p_i)/t$ and hence the spacetime does not admit an US. On the other hand, if two $p_i$'s are equal it follows that $p_2=p_3=-2p_1=2/3$ (without loss of generality). Then the line element represents a Petrov type $D$, non-stationary, plane-symmetric vacuum which, according to theorem \ref{Theorem 1}, admits a one-degree freedom of USs (cf. the end of this section).\\ 

In theorem 3 of \cite{Barnes1} it is claimed that all vacuum
spacetimes admitting an US are static, which
would generalize Tr\"{u}mper's result~\cite{Trumper1} by including Petrov type
$D$. However, this conclusion only holds when $\mu\rho\geq 0$. Indeed,
a static member of ${\cal D}_0$ necessarily admits a rigid (i.e.\
non-expanding) US, such that $\mu\rho\geq 0$, cf.\ (\ref{mu=qrhoc}).
Conversely, when $\mu\rho=0$ or $\mu\rho>0$ for a PE member, it
admits the HO timelike KVF (\ref{KVF x4}), resp.\ (\ref{KVF psi}) and is thus static. This is in agreement with proposition \ref{Prop: static}: regarding
$\mu=\rho=0$ criterion 6'' tells that in fact all boost-isotropic spacetimes, with
$\pi=-\tauc$ wrt a WPNT, are static, while for $\mu\rho>0$ one checks that criterion 2'' is satisfied by virtue of  (\ref{Psis})-(\ref{def FG}).
In appendix B the freedom of the rigid USs and HO timelike KVF directions (static observers)
in these cases is also specified, which is in accordance with a
result by Wahlquist and Estabrook~\cite{WahlquistEstabrook}. In summary we have:
\begin{Theorem}\label{Theorem 2} A Petrov type $D$ Einstein space is static if it admits a
rigid US. This is precisely the case when the spacetime is PE and has a positive or zero
scalar invariant $\mu\rho$, being the product
of the divergences of (non-twisting) Weyl principal null vectors $k^{a}$ and $l^{a}$ subject to $k^al_a=-1$.
For $\mu\rho>0$ there
is an up to reflection unique rigid US, defined from the geometry by
(\ref{udef}) and $q=\mu/\rho$ and parallel to the unique
HO timelike KVF direction. For $\mu\rho=0 \Rightarrow \mu=\rho=0$
(classes IA and IC) all USs are rigid USs and have a one-degree
freedom, while the HO timelike KVF directions are parametrized by two
constants.
\end{Theorem}

For completeness we display standard coordinate forms of the PE Petrov type D Einstein space
metrics, as recovered here by (\ref{ds2HO}), (\ref{def Psi}) and
(\ref{ds2perp constcurv}), (\ref{ds2Sigma constcurv}), (\ref{ds2perp
nonconst}), (\ref{ds2Sigma nonconst}).

$C^{00}$ corresponds to (\ref{ds2perp constcurv}) and (\ref{ds2Sigma constcurv}). From
 (\ref{tau=0 conseq}), (\ref{Kperp}), (\ref{mu=0conseq}) and (\ref{KSigma}) one deduces that
\begin{eqnarray*}
K^\perp=-3x=-3y=K^\Sigma,\quad \Psi_2=-\frac{\L}{3}=4x^3\neq 0.
\end{eqnarray*}
Rescaling
$\zeta$, $u$ and $v$ by a factor $(2x)^{-1}$  one arrives at
\begin{equation*}
\di s^2=\frac{2\di
\zeta\di\bar{\zeta}}{1+\frac{\L}{2}\zeta\bar{\zeta}}-\frac{2\di
u\di v}{1-\frac{\L}{2}uv},\quad 1+\frac{\L}{2}\zeta\bar{\zeta}>0,\quad \L\neq 0.\label{C00}
\end{equation*}
This represents the Einstein space limit $\Phi_0=0$ of
Bertotti's static and homogeneous electrovac family with cosmological
constant~\cite{Bertotti,Robinson}, exhibiting spatial rotation and
boost isotropy (complete group $G_6$ of isometries). The $\L=0$ limit yields flat Minkowski
spacetime.

$C^0{}_+$ and $C^0{}_-$
correspond to (\ref{ds2perp constcurv}) and (\ref{ds2Sigma nonconst}), resp.\  (\ref{ds2Sigma constcurv}) and (\ref{ds2perp
nonconst}). Making use of
(\ref{taumurhospec}), replacing in the $C^0{}_+$ ($C^0{}_-$) case the
coordinate $y$ ($x$) by $r=-(2m)^{1/3}/(x+y)$, rescaling the
remaining coordinates by a factor $(2m)^{-1/3}$ and writing $Y\equiv(2m)^{1/3}>0$ one finds
\begin{eqnarray*}
&&\di s^2=r^2\left(\di \xi^2+\delta
\cos(\sqrt{k}\,\xi)^2\di\eta^2\right)+\frac{\di r^2}{g_k(r)}-\delta g_k(r)\di \chi^2,\nonumber\\
&&g_k(r)=k-\frac{2m}{r}-\frac{\L}{3}r^2,\quad k=K(2m)^{1/3}
\in\{-1,0,1\},
\label{grk}
\end{eqnarray*}
with $\d=1$ for $C^0{}_+$ ($\d=-1$ for $C^0{}_-$). 
These solutions have a complete group
$G_4$ of isometries acting on spacelike (timelike) three-dimensional orbits, and for $\L=0$ correspond to Kinnersley's case I (IV) with $l=0$.
The static region of $C^0{}_+$ ($g_k(r)$>0) yields class A in the classification of static Petrov type $D$ vacua by Ehlers
and Kundt \cite{EhlersKundt}; $C^0{}_-$ is static everywhere and corresponds to class B.
Regarding $C^0_+$, the subcase $k=1$ reproduces after $\xi\mapsto \pi/2-\xi$ the
well-known forms of the spherically symmetric
Schwarzschild-Kottler interior and exterior metrics~\cite{Schw,Kottler}; the subcase $k=\L=0$, $r>0$ ($g_k(r)<0$) gives another form of the plane-symmetric Kasner metrics (cf.\ supra).

$C^*$ corresponds to (\ref{ds2perp nonconst}) and (\ref{ds2Sigma nonconst}), which gives the line element
\begin{eqnarray}
&&\di s^2=
\frac{1}{(x+y)^2}\left(\frac{\di
x^2}{f(x)}+f(x)\di\phi^2+\frac{\di
y^2}{g(y)}-g(y)\di\psi^2\right),\nonumber\\
&&f(x)=x^3+Cx+D>0,\quad g(y)=-f(-y)-\frac{\L}{3}.\label{C*}
\end{eqnarray}
The KVFs $\partial_\phi{}^a$ and $\partial_\psi{}^a$
generate the complete, abelian group $G_2$ of isometries.
For $\L=0$, (\ref{C*}) is the form of the C-metric
obtained by Levi-Civita and recovered by Ehlers and Kundt, and corresponds to Kinnersley's case IIIA. It is generally assumed - and suggested in the original paper \cite{KinnersleyWalker} - that the Kinnersley-Walker form
\begin{eqnarray}
&&\di s^2=
\frac{1}{\a^2(\xi+\eta)^2}
\left(\frac{\di
\xi^2}{h(\xi)}+h(\xi)\di \phi^2+\frac{\di
\eta^2}{k(\eta)}-k(\eta)\di \psi^2\right),\nonumber\\
&&h(\xi)=1-\xi^2-2m\alpha \xi^3>0,\quad k(\eta)=-h(-\eta)\label{C*Kinn}
\end{eqnarray}
equivalently describes the gravitational field of the $\L=0$ C-metric. However, this is not entirely correct.
Equating the Lorentz invariants appearing in the right hand sides of the equations in (\ref{def xy})-(\ref{def D}), calculated for the metrics (\ref{C*}) and (\ref{C*Kinn}), yields
\begin{eqnarray}
&&x=-(2m)^{\frac{1}{3}}\left(\a\xi+\frac{1}{6m}\right),\, y=-(2m)^{\frac{1}{3}}\left(\a\eta-\frac{1}{6m}\right),\nonumber\\
&&C=-\frac{1}{3(2m)^{\frac{4}{3}}},\quad D=\a^2-\frac{1}{54m^2}.\label{CD}
\end{eqnarray}
Hence (\ref{C*Kinn}) only covers the range $C<0$, $D>-2\left(-\frac{C}{3}\right)^{\frac{3}{2}}$, whereas in general the constant scalar invariants $C$ and $D$ are allowed to take any real value. Yet, the cubic $f(x)$ has discriminant $-4C^3-27D^2$; thus it has three distinct real roots if and only if
\begin{equation}\label{CD cond}
C<-3\left(\frac{D}{2}\right)^{\frac{2}{3}}\quad \Leftrightarrow \quad
C<0,\, |D|<2\left(\frac{-C}{3}\right)^{\frac{3}{2}}.
\end{equation}
Thus (\ref{CD}) {\em is} compatible for this case, and by further rescaling $\phi$ and $\psi$ with a factor $\a(2m)^{-1/3}$ one arrives at (\ref{C*Kinn}); (\ref{CD cond}) is equivalent with $m\a<1/\sqrt{27}$, leading to the physical interpretation of two uniformly accelerating masses.
Recently, Hong and
Teo~\cite{HongTeo1} introduced a normalized factored form for this situation, which greatly simplifies certain analyses of the
C-metric. A further coordinate transformation can be made such that
the Schwarzschild metric is comprised as the subcase $\alpha=0$. This
was further exploited for the full {\cal D}-class in \cite{Griffiths2}.

Finally, we write down the equations which determine all USs for a member of $C^*$, in the coordinates $y$ and $\psi$ of (\ref{C*}). Let
\begin{equation*}
U^a=-\frac{x+y}{\sqrt{g(y)}}\partial_t{}^a,\quad V^a=-(x+y)\sqrt{g(y)}\partial_y{}^a
\end{equation*}
in the static region and
\begin{equation*}
U^a=-(x+y)\sqrt{-g(y)}\partial_y{}^a,\quad V^a=\frac{x+y}{\sqrt{-g(y)}}\partial_t{}^a
\end{equation*}
in the non-static region, and gauge-fix $k^a=(U^a+V^a)/\sqrt(2)$, $l^a=(U^a-V^a)/\sqrt(2)$. The unit timelike field (\ref{udef}) is an US if and only (\ref{deriv q}) holds; this translates to $q=q(y,\psi)$ and
\begin{eqnarray}\label{US C}
g(y)(q\pm 1)q_{,y} + (q\mp 1)(q_{,\psi}+g'(y)q)=0.
\end{eqnarray}
Here and below the upper (lower) signs should be taken in the static (non-static) region. For solutions $q=q(y)$, i.e.\ $q_{,\psi}=0$, direct integration of (\ref{US C}) yields
\begin{equation}\label{q(y)}
g(y)(q(y)\mp 1)^2=E_{\pm} q(y),
\end{equation}
with $E_{+}>0$ and $E_{-}<0$ constants of integration.
Notice that in the static region the solution $q(y)=1$ yields the unique static observer.  In the case $q_{,\psi}\neq 0$ the solutions get implicitly determined by an equation of the form $\psi=\psi(y,q)$, on applying the method of characteristics for first-order PDEs (see e.g.\ \cite{Pinchover}). In the subcase where $C=D=\L=0$, $g(y)$ reduces to $y^3$ and this equation reads
\begin{equation}
\psi=-\left(\frac{y(q\mp1)^{\frac{2}{3}}}{q^{\frac{1}{3}}}\right)^2\int \frac{(q\mp1)^{\frac{4}{3}}\di q}{3q^{\frac{5}{3}}}+Z\left(\frac{y(q\mp1)^{\frac{2}{3}}}{q^{\frac{1}{3}}}\right).\label{eC}
\end{equation}
Here $Z$ is a free function of its argument, making the one-degree freedom of USs more explicit.
Replacing (\ref{ds2perp nonconst}) by (\ref{ds2perp constcurv}) does not alter these equations, i.e., the above remains valid for $C^0{}_+$. Then $C=D=0$ is equivalent to $x=K^\bot=0$, cf.\ (\ref{tau=0 conseq}) and (\ref{Kperp}); for $\L=0$ the non-static region $y<0$ corresponds to the plane-symmetric Kasner vacuum metrics, where $y=-(3t/2)^{-2/3}$ and a rescaling of the other coordinates recovers (\ref{Kasnermetric}), the USs being determined by (\ref{q(y)}) and (\ref{eC}) with the lower sign.

\section{Perfect fluid generalizations of the
C-metric}\label{section:pfmetric}

\subsection{Line element}\label{subsection: ID line element}

We resume the integration of class I started at the end of $\S$~\ref{subsection:Einstein spaces}.
We thereby focus on the subclass
ID characterized by $\tau\neq0\neq \mu\rho$. Let us first summarize
what we did so far. We started off with the closed set (\ref{deriv
q})-(\ref{def FG}) of first-order GHP equations in the seven
(weighted) real variables $\P$, $S$, $w$, $\mu$, $\rho$, $q$,
$\eth'\tau$ and the complex variable $\tau$. These variables are
equivalent to two dimensionless spin and boost gauge fields, e.g.\
$\tau/\tauc$ and $\mu/\rho$, and seven real scalar invariants. The
boost and spin gauge fields could serve to invariantly fix the
tetrad -- the ID members being therefore anisotropic -- but can be
further ignored. For the $C^*$-Einstein spaces, $S=0$ and
$w=\Lambda=const$, and we remarked that $q$ is not a part of the
intrinsic describing set of variables. Hence we end up with four
real scalar invariants in this subcase. These invariants are
equivalent to the two constants $C$ and $D$ and two independent
functions $x$ and $y$, which we took as coordinates and in terms of
which, on adding two coordinates $\phi$ and $\psi$ related to the
symmetries, the corresponding $C^*$-metric can be expressed. In the
perfect fluid case $S\neq 0$, (\ref{Phi_ii}) gives the boost field $q$ which, starting from an arbitrary gauge $(k^a,l^a)$, turns $u^a$ given by (\ref{udef}) into the invariantly-defined fluid four-velocity.
The four invariants and their use
persist, just as the coordinate $\phi$. However,
the scalar invariants $w$ and $S$ are no
longer constant and $\psi$ is no longer
a suitable coordinate. Thus we
need one more scalar invariant for our description and one remaining
coordinate complementary to $y$.

For the first purpose it is natural to look at the kinematics of the
fluid, which are fully determined by
\begin{eqnarray}
&&\bu \equiv 2\nabla_{(a} u_{c)} m^{a}\mc^{c} =\nabla_a u_c
v^av^b=\frac{\theta}{3},\label{def bu}\\
&&\upar\equiv v_a\dot{u}^a,\quad \dot{u}^\perp_a\equiv
2\mc_{(a}m_{c)}\dot{u}^c.
\end{eqnarray}
Here
\begin{equation}\label{vdef}
v^a\equiv\frac{1}{\sqrt{2q}}\left(qk^a-l^a\right)
\end{equation}
is the intrinsic spacelike vector field, which
determines at each point the up to
reflection unique normalized vector orthogonal to $u^a$ and lying in the PND plane $\Sigma$,
while $\upar$ and $\dot{u}^\perp_a$ are the component along $v^a$,
resp.\ projection onto $\Sigma^\perp$ of the acceleration
$\dot{u}^a$. In analogy with (\ref{def bu}) we define the invariant
\begin{eqnarray}
\bv \equiv 2\nabla_{(a} v_{c)} m^{a}\mc^{c}.\label{def bv}
\end{eqnarray}
The relation with GHP quantities is
\begin{eqnarray}
&&\bu=\frac{\mu-q\rho}{\sqrt{2q}},\quad
\bv=-\frac{\mu+q\rho}{\sqrt{2q}},\label{def bu bv}\\
&&\upar=(2q)^{-3/2}(\tho'q+q\tho q) =\frac{\tho q}{\sqrt{2q}}-\bu,\\
&&-\dot{u}^\perp_a=\tauc m_a+\tau \mc_a\equiv \alpha_a=\frac{\di_a
x}{x+y}.\label{defalphaa}
\end{eqnarray}
Notice that (\ref{def bu bv}) is equivalent to
\begin{eqnarray}
\bu\,u_a-\bv\,v_a= \mu k_a-\rho l_a\equiv \beta_a=\frac{\di_a
y}{x+y}.\label{defbetaa}
\end{eqnarray}
In combination with (\ref{def F})-(\ref{def G}), (\ref{defalphaa}) and (\ref{defbetaa}) imply
\begin{eqnarray}
&&2\tau\tauc=\dot{u}^\perp_a\dot{u}^{\perp a}=x^3+Cx+D,\\
&&2\mu\rho=\bv^2-\bu^2=y^3+Cy-D-\frac{w}{3}.\label{bu^2bv^2 rel}
\end{eqnarray}
We choose $\bu$ as the final describing invariant and use $\bv$ and
$\upar$ as auxiliary variables. In view of (\ref{def bu
bv})-(\ref{defalphaa}) one deduces that  the differential
information for $S$, $w$ and $\bu$ comprised in (\ref{deriv
q})-(\ref{def FG})
is precisely
\begin{eqnarray}
&&D_a S=-S\dot{u}_a, \label{DS}\\
&&\di_a w=-\u(w)u_a,\quad \u(w)=-3\bu S,\label{dw}\\
&&\di_a \bu=-\u(\bu)u_a,\quad
\u(\bu)=-\v(\bv)+\bv(\upar-\bv)-\frac{S}{2},\;\;\;\;\;\;\;\;\label{dbu}\\
&&\v(\bv)=-\frac{x+y}{2}(3y^2+C).\label{dbv}
\end{eqnarray}
From (\ref{dbv}) it follows that $\bv$ is non-constant, such
that we may see the second part of (\ref{dbu}) as a definition of
$\upar$. (\ref{dw}) is nothing but the energy
resp.\ momentum conservation equations for a perfect fluid subject
to $D_a w=0$. The first part of equation (\ref{dbu}) confirms that
$D_a\t=0$ ~\cite{Trumper,Barnes1}, whilst the second implies again
that the expansion scalar does not vanish in general
(cf.\ the end of $\S$~\ref{subsection:Integrability} and below).

For the second purpose we rely on the hypersurface-orthogonality of
$u^a$
by assumption: zero-weighted real scalar fields $t$ and $I$ exist
such that
\begin{equation}
\di_a t=Iu_a.\label{dt}
\end{equation}
The integrability condition hereof is
\begin{equation}
D_a I=-I\dot{u}_a=-I(\dot{u}^\perp_a+\upar v_a),\label{DI}
\end{equation}
which is equivalent to
\begin{equation}
\eth I=\tau I,\quad \v(I)=-\upar I.\label{eth v I}
\end{equation}
From $\bv\neq 0$, (\ref{defbetaa}) and
(\ref{dt}) it follows that $t$ is functionally
independent of $y$ (and of $x$ and $\phi$) and we take it as the fourth coordinate.
With the aid of (\ref{dt})-(\ref{DI}), (\ref{DS}) and the first parts of (\ref{dw}) and (\ref{dbu}) precisely tell that $A\equiv \frac{S}{2I}$, $w$ and $\bu$ only depend on $t$.
Hence $\bv=\bv(y,t)$ from (\ref{bu^2bv^2 rel}). On using (\ref{defalphaa})-(\ref{defbetaa}), the first part of
(\ref{eth v I}) is equivalent to $J=J(y,t)$, where $J\equiv \frac{x+y}{I\bv}$.
Eliminating
$\upar$ between the second parts of (\ref{dbu}) and (\ref{eth v I}), and using $\v(x+y)=-\bv(x+y)$ implied by (\ref{defalphaa})-(\ref{defbetaa}), yields
\begin{eqnarray}
\bv^2\v(J)=\bv J\u(\bu)+A(x+y).\label{vT}
\end{eqnarray}
Inverting (\ref{defbetaa}) and
(\ref{dt}) we get
\begin{eqnarray}
(x+y)u_a=\bv J\di_a t,\quad (x+y)v_a=\bu J\di_a t-\frac{\di_a
y}{\bv},\label{uv covar t}
\end{eqnarray}
or dually
\begin{equation}
-\frac{u^a}{x+y}=\frac{\partial_{t}{}^a}{\bv J}+\bu\,\partial_y{}^a,
\quad -\frac{v^a}{x+y}=\bv\,\partial_y{}^a.\label{uv contravar t}
\end{equation}
Thus in the chosen coordinates (\ref{vT}) reads
\begin{equation}\label{Jy}
\partial_y J(y,t)={\bv(y,t)^{-3}}\left[\bu'(t)-A(t)\right].
\end{equation}
From (\ref{ds2HO}), (\ref{ds2perp nonconst}), $g^\Sigma_{ab}=(v_av_b-u_au_b)/(x+y)^2$ and the only remaining equation (\ref{dw}) we obtain the line element
\begin{eqnarray}
&&\di s^2=(x+y)^{-2}[\di s_{\perp}^2+\di s_{\Sigma}^2],\label{ds2 t}\\
&&\di s_\perp^2=\frac{\di x^2}{2\tau\tauc}+2\tau\tauc\di \phi^2,\quad
2\tau\tauc=x^3+Cx+D,\;\;\;\label{diCperp}\\
&&\di s_{\Sigma}^2 =\left(\bu J\di t-\frac{\di
y}{\bv}\right)^2-\left(\bv J\di t\right)^2,\label{ds2Sigma t}
\end{eqnarray}
where
\begin{eqnarray}
&&\bu=\bu(t),\quad w=w(t),\quad A\equiv\frac{\bv J S}{2(x+y)}=A(t),\label{wt but S/I}\\
&&w'(t)=6\bu(t)A(t)\, \Leftrightarrow\,\di_a w = 6\bu~A\,\di_a t=3\bu\,Su_a,\label{dw/dt}\\
&&\bv=\bv(y,t)=\sqrt{y^3+Cy-D+\bu(t)^2-w(t)/3},\label{bv t}\\
&&J=J(y,t)=\left[\bu'(t)-A(t)\right]\int \frac{\di
y}{\bv(y,t)^{3}}+L(t), \label{J expr t}\;\;\;\;\;\;\;\;\;\;\;\;
\end{eqnarray}
with $L(t)$ a free
function of integration.
The solutions are defined and regular in the coordinate regions
\begin{eqnarray}
\label{coord range x}&&2\tau\tauc\equiv x^3+Cx+D>0,\\
\label{coord range y}&&\bv(y,t)^2\equiv y^3+Cy-D+\bu(t)^2-w(t)/3>0.\;\;\;\;\;\;\;\;\;\;\;\;
\end{eqnarray}

Notice that we nowhere used $S\neq 0$
explicitly in the above integration procedure. Therefore, the above line
element describes the {\em complete} class ID, including the
$C^*$-vacuum limits which correspond to $w(t)=\L$ and $A(t)=0$, cf.\ (\ref{wt but S/I}). In this case the coordinate transformation $(t,y,x,\phi) \mapsto (\psi,y,x,\phi)$, which connects (\ref{ds2 t})-(\ref{J expr t}) to the original form (\ref{C*}), eliminates $\bu(t)$ and $L(t)$ and follows from (\ref{udef}), (\ref{KVF psi}),  (\ref{vdef}), (\ref{def bu bv}) and (\ref{uv covar t}), giving
\begin{eqnarray}
\di_a\psi & = & -\frac{x+y}{2\mu\rho}\left(\mu k_a+\rho l_a\right)= \frac{x+y}{2\mu\rho}\left(\bv u_a-\bu v_a\right) \nonumber\\
& = &  J\di_at+\frac{\bu}{\bv\left(\bv^2-\bu^2\right)}\di_ay.
\end{eqnarray}
Hence, $\psi=\psi(y,t)$ and it is the solution of the consistent system
\begin{equation}
\partial_t\psi=J, \quad \partial_y\psi = \frac{b}{\bv\left(\bv^2-b^2\right)},
\end{equation}
the integrability condition hereof being precisely (\ref{Jy}) with $A(t)=0$.
The transformation is singular at  degenerate roots of $\bv^2$ and at the union of the black hole and acceleration horizons~\cite{Griffiths,Griffiths2} $\bv^2-b^2\equiv f(-y)+\L/3=0$, which separate the static from the non-static regions. Let us emphasize that the $\bu(t)$-freedom is essentially a freedom in the choice of coordinates. The form (\ref{C*}) describes the full C-metric manifold; $y$ can take any value, and the sign of $f(-y)+\frac{\Lambda}{3}$
is positive in the static region and negative in the non-static region. In the form (\ref{ds2 t})-(\ref{ds2Sigma t}) $y$ is always spacelike and 
we have constructed $t$ as a synchronized timelike coordinate corresponding to an US $u^a$, with associated expansion rate $\t(t)=3\bu(t)$; for fixed $\bu(t)$ the range of $y$ is constrained by (\ref{coord range y}) and only
this subregion of the manifold is described by the coordinates. E.g.\ (\ref{ds2 t})-(\ref{J expr t}) with $A(t)=0$, $w(t)=\L/3$, $\bu(t)=0$ and $L(t)=1$ (which formally reduces to (\ref{C*}) on putting $t=\psi$ additionally) only describes the static part of the C-metric, the vector field $u^a$ then lying along the unique HO timelike KVF direction. However, in the neighborhood of any point with coordinate label $y$, the metric can be described by (\ref{ds2 t})-(\ref{J expr t}), by choosing $\bu(t)^2>f(-y)+\L/3$.

We neither used $\tau\neq 0$. This implies that the line element of
the complete class IB, characterized by $\mu\rho\neq 0 =\tau$ and
constituted by all LRS II Einstein spaces and shear-free perfect
fluids with $D_a w=0$, is described by (\ref{ds2 t})-(\ref{J expr t}), with (\ref{diCperp}) replaced by (\ref{ds2perp constcurv}). This class
was first described by
Kustaanheimo~\cite{Kustaanheimo} and rediscovered by Barnes \cite{Barnes1}, both using different coordinates (see also (16.49), (16.51) in \cite{SKMHH}).

Of course, the result (\ref{ds2 t})-(\ref{J expr t}) could have been obtained without referring to GHP calculus.
Barnes~\cite{Barnes1} showed that the metric can be written in the form
\begin{eqnarray}\label{BarnesMetr}
\di s^2=(x+Y)^{-2}{\left(f^{-1}\di x^2+f\di\phi^2 + \di
z^2-e^{2Z}\di t^2\right)},\;\;\;\;\;
\end{eqnarray}
with $f=f(x)$. Indeed, 
from (\ref{uv covar t}), (\ref{Jy}) and (\ref{bv t}) it follows that $(x+y)v_a$ is exact:
$(x+y)v_a=\di_a z$; $z$ is used as a coordinate instead of
$y$, and one puts $J\bv\equiv e^{Z}$, $Z=Z(y,t)$. Notice from (\ref{defbetaa}) that now
\begin{eqnarray}
y=Y(z,t),\quad Y_{,z}=-\bv,\quad \theta=3Y_{,t}e^{-Z}.
\end{eqnarray}
Let us directly attack the field equations in these coordinates, thereby correcting \cite{Barnes1}.
One can check that only four of the field equations are not identically
satisfied (the indices 1 to 4 label the Weyl principal tetrad vectors naturally associated with (\ref{BarnesMetr})):
\begin{eqnarray}
&&G_{34}=0=  -Y_{,tz}+Y_{,t} Z_{,z},\label{G34}\\
&&G_{11}-G_{33} =0 =\label{G1133}\\ &&\;\;\;f' -2 Y_{,zz}
+(x+Y)\left(Z_{,z}^2+Z_{,zz} - f''/2
\right),\label{G1133}\nonumber\\
&&G_{11} = p=\label{G11}\\
&&\;\;\;
2\left(e^{-2Z}\left(Y_{,tt}-Y_{,t}Z_{,t}\right)-Y_{,z}Z_{,z}
-f'/2
\right)\left(x+Y\right)\nonumber\\
&&\;\;\;+\left(Z_{,zz}+Z_{,z}^2\right)\left(x+Y\right)^2+3\left(Y_{,z}^2+f-Y_{,t}^2e^{-2Z}\right),\nonumber \\
&&G_{33}+G_{44} = S=\label{G3344}\\
&&\;\;\;2\left(x+Y\right)\left[Y_{,zz}-Y_{,z}Z_{,z}\right.
+\left.e^{-2Z}\left(Y_{,tt}-Y_{,t}Z_{,t}
\right)\right].\nonumber
\end{eqnarray}
Hence, \emph{if supplemented with $\theta\sim Y_{,t}=0$}, these
equations are the ones obtained in Barnes~\cite{Barnes1}: equation
(\ref{G34})~$\equiv \v(\t)=0$ was missed out, and both equations
(\ref{G11}) and (\ref{G3344}) differ from equations (4.23), resp.\
(4.24) in~\cite{Barnes1} by a term $2(x+Y)Y_{,tt}e^{-2Z}$. Thus, it
is clear that with these differences a correct non-expanding
solution can be found, but the analysis of expanding solutions will
be incorrect.

Differentiating (\ref{G1133}) twice wrt $x$ yields
${d^4f(x)}/{dx^4}=0$, whence
\begin{equation}
f(x)=ax^3+bx^2+cx+d.
\end{equation}
Substituting this in equation (\ref{G1133}), and equating
coefficients of powers of $x$, leads to
\begin{eqnarray}
&&Z_{,zz}(z,t) + Z_{,z}(z,t)^2\,=\,3aY(z,t)-b,\\
&&Y_{,zz}(z,t)\,=\,\frac{c}{2}-Y(z,t)b+\frac{3}{2}a Y(z,t)^2.
\end{eqnarray}
The solutions $Y(z,t)$ of the last equation are defined by 
\begin{equation}
\int^{Y(z,t)} \frac{dr}{\sqrt{ar^3-br^2+cr+f_1(t)}}-z+f_2(t)=0,
\end{equation}
which can be solved for $z$ in terms of $Y$. This eventually suggests to transform coordinates from $(z,t)$ into
$(y,t)$, with $y=Y(z,t)$. Rescaling and translating coordinates
allows us to set $a=1$ and $b=0$. One can check that the remaining
equations lead exactly to equations (\ref{Jy}) and (\ref{dw/dt}),
recovering solution (\ref{ds2 t})-(\ref{J expr t}).

\subsection{Properties}\label{subsection: Properties}

Consider the metric (\ref{ds2 t})-(\ref{J expr t}), for which we assume henceforth that it describes a perfect fluid ($A(t)\neq 0$).
In contrast to the Einstein subcase, $u^a$ is now the unique invariantly-defined fluid velocity, and the expansion rate $\t(t)$ and energy density $w(t)-\L$ of the fluid are scalar invariants. Expressions for the pressure $p+\L$ and the components $\dot{u}^{\bot a}$ and $\upar$ of the acceleration follow from (\ref{defalphaa}), (\ref{dbu})-(\ref{dbv}), (\ref{uv contravar t}) and (\ref{wt but S/I}):
\begin{eqnarray*}
&&p=\frac{2(x+y)A(t)}{(\bv J)(y,t)}-w(t),\quad\dot{u}^{\bot a}=-(x+y)f(x)\partial_x{}^a,\\
&&\upar=\bv(y,t)-\frac{x+y}{\bv(y,t)}\left(\frac{3y^2+C}{2}+\frac{b'(t)-A(t)}{(\bv J)(y,t)}\right).
\end{eqnarray*}
The fluid is non-shearing and non-rotating, i.e.\ $u^a$ is an US. Because of (\ref{geod shearfree})-(\ref{pitau}) criterion 5 of proposition \ref{Prop: US} is satisfied, such that there is a one-degree freedom of USs.
These can be found by taking $q=1$ in (\ref{udef}) and (\ref{vdef}), hereby fixing the ($k^a,\,l^a$)-gauge geometrically, and solving  $(\ref{deriv q})$ with $q$ replaced by $Q$,
the USs then being $(Qk^a+l^a)/\sqrt{2Q}$. This yields $Q=Q(y,t)$ and
\begin{eqnarray*}
&&\bv J[\bv(Q+1)+\bu(Q-1)]Q_{,y}+(Q-1)Q_{,t}\\&&=-2Q(Q-1)\bv(\bv J)_{,y}\\
&&=-\frac{Q(Q-1)}{\bv}\left[(3y^2+C)\bv J+2(\bu'-A)\right].
\end{eqnarray*}

If the class is to be used as a cosmological model, it is interesting to discuss the intrinsic freedom. By (\ref{dw/dt})
and (\ref{uv covar t}) we have that $2A(t)\di_a t=S u_a$ and
$J(y,t)\di_a t=(x+y)u_a/\bv$ are invariantly-defined one-forms, and
hence so is $L(t)\di_a t$ because of (\ref{J expr t}).
It follows
that $\frac{L}{A}(t)$ is a scalar invariant. Moreover, as $A(t)\di_a
t$ is exact we may remove the only remaining coordinate freedom on
$t$ by putting $A(t)=1$, such that the conservation of energy
equation (\ref{dw/dt}) can be considered as a definition
$\theta(t)=w'(t)/2$. Hence, in this most
general picture for $S\neq 0$, {\em the scalar constants
$C$, $D$ and invariants $\frac{L}{A}(t)$, $w(t)$ characterize the model within the class.}
Notice that the presence of two invariantly-defined, distinguishing
free functions could have been predicted, since after elimination of
$\upar$, there are two scalar invariants $\u(\bu)$ and
$\u(S)$ remaining unprescribed in the system of equations (\ref{DS})-(\ref{dbv}).

In this fashion however, the physical implications remain obscure:
it would be nice to have a free function, with a clear physical interpretation, instead of
$L/A$. Spacetimes with $L(t)=0$ have $w(t)$ as the only free
function. If $L(t)\neq 0$, $L(t)$ can alternatively be fixed to 1 by a $t$-coordinate transformation.
In this case the metric structure functions
display the expansion scalar, the energy
density and the pressure (since $2A(t)\di_a t = (w+p)u_a$);
these are related by energy conservation (\ref{dw/dt}), where $w(t)$ and $A(t)$
can be chosen freely. Alternatively, one can
subdivide further in $\theta=0$ and $\theta \neq 0$. In the case
$\theta=0$, the energy density $w-\L$ is constant because of (\ref{dw/dt})
and can be chosen freely, just as $A(t)$. In the most interesting
case $\theta\neq 0$, $w(t)$ and $\theta(t)$ can be chosen freely,
determining $S u_a$ via (\ref{dw/dt}). Thus {\em class ID provides a class of
anisotropic cosmological models with arbitrary evolution of matter
density and (non-zero) expansion.}

Regarding symmetry, all perfect fluid ID models admit at least one
KVF $\partial_\phi{}^a$ given by (\ref{KVF phi}), which at each
point yields an invariantly-defined spacelike vector orthogonal
to $\dot{u}^{\perp a}$ and lying in $\Sigma^\perp$. If $\phi$ is
chosen to be a periodic coordinate, with range given by $\left[-\pi
E, \pi E\right[$, the spacetime is cyclically symmetric. We will
then refer to the region $F\equiv f(x)=0$, where the norm of
$\partial_{\phi}{}^a$ vanishes, as the axis of
symmetry~\cite{Griffiths}~\footnote {Strictly speaking, with the
terminology of \cite{SKMHH}, the spacetime has in general only a
cyclic symmetry, as the axis will be shown to be irregular, and
consequently not part of the spacetime.}. Finding the complete group
of isometries and their nature is trivial in our approach. The
functions $x$, $y$, $w$ and $L/A$ are invariant scalars, such that
$K^a\di_a x=K^a\di_a y=K^a\di_a w=K^a\di_a \frac{L}{A}=0$ for any
KVF $K^a$. As the ID models are anisotropic, it follows that the
complete isometry group is at most $G_2$, and if it is $G_2$, both
$w$ and $L/A$ are constant. Conversely, when $w$ and $L/A$ are
constant we have $\t\equiv 3b=0$ from (\ref{dw/dt}), $\bv=\bv(y)$
from (\ref{bv t}) and $J(y,t)=-A(t)F_2(y)$ from (\ref{J expr t}). By
redefining the time coordinate such that $A(t)=1$ one sees from
(\ref{ds2 t})-(\ref{J expr t}) that $\partial_t{}^a$ is a HO
timelike KVF. We conclude that {\em the ID perfect fluid models have
at least one spacelike KVF $\partial_\phi{}^a$, which may be
interpreted as the generator of cyclic symmetry. They admit a second
independent KVF if and only if both scalar invariants $w$ and $L/A$
are constant, in which case the spacetimes are static and the
complete group of isometries is abelian $G_2$, generated by
$\partial_\phi{}^a$ and $\partial_t{}^a$.}

Consider the case where $f(x)$ has 3 real non-degenerate roots $x_i$,
i.e.\ (\ref{CD cond}) holds. If $x_1<x_2<x_3$ then
$f(x) > 0$ for all $x\in \left]x_1,x_2\right[$. Furthermore, we let
$\phi$ be a periodic coordinate. The ratio between circumference and
radius of a small circle around the axis, $x=x_1$ or $x=x_2$, is
given by
\begin{equation}
\lim_{\stackrel{x\rightarrow x_2}{<}}\frac{2\pi E
\sqrt{f(x)}}{\int_{x}^{x_2}\sqrt{f^{-1}(x)}\di x} = -\pi E
\left(3x_2^2+C\right)
\end{equation}
respectively
\begin{equation}
\lim_{\stackrel{x\rightarrow x_1}{>}}\frac{2\pi E
\sqrt{f(x)}}{\int_{x_1}^{x}\sqrt{f^{-1}(x)}\di x} = \pi E
\left(3x_1^2+C\right).
\end{equation}
It is only possible to choose the parameter $E$ such that the
complete axis is regular, if  $3x_1^2+C=-(3x_2^2+C)$. However,
eliminating $C$ and $D$ between this equation and $f(x_1)=f(x_2)=0$
implies $x_1=x_2$. Consequently, if $f(x)$ has three real
non-degenerate roots, the spacetime contains a conical singularity.
This echoes the properties of the
C-metric~\cite{KinnersleyWalker,Griffiths}, and suggests the presence of a cosmic string.


\section{Conclusions and discussion}

A new class of Petrov type $D$ exact solutions of Einstein's field equation
in a perfect fluid with spatially homogeneous energy density has been presented. It consists of all anisotropic such fluids with shearfree normal four-velocity, and
generalizes a previously found class to include non-zero expansion.
The analysis and integration was rooted in the 2+2 structure of the metric and use of invariant quantities.
This approach clarified the link with the vacuum C-metric limit, and certain
properties of the vacuum case are inherited. However, the presence of the perfect fluid
defines generically two extra invariants. For the expanding solutions, this translates into an evolution of energy density and expansion which can be chosen freely. This subclass contains only one (potentially cyclic) symmetry.

The viability of these
solutions as a low-symmetry class of cosmological models is subject to further research.
More in particular, it should be clarified whether a thermodynamic
interpretation of the perfect fluid can be made \cite{Coll} - it is
certainly not possible to prescribe a barotropic equation of state $p=p(w)$.
The relation with the C-metric also suggests to further examine
the arising coordinate ranges, properties of 
horizons, and whether an interpretation as a perturbation for small masses
of a known PF solution exists for certain members.

\begin{acknowledgments}
The authors would like to thank N.\ Van den Bergh and S.~B.\ Edgar for reading the document and useful suggestions.
\end{acknowledgments}

\appendix
\section{Geroch-Held-Penrose (GHP) formalism}

The GHP formalism~\cite{GHP,SKMHH} is a complex, scalar formalism, which is a `weighted' version of the Newman-Penrose (NP) tetrad
formalism. Use is made of a complex null tetrad
$(\ee_1{}^a,\ee_2{}^a,\ee_3{}^a,\ee_4{}^a)\equiv\left(m^a,\overline{m}^a,l^a,k^a\right)$, where
\begin{equation}\label{normal conds}
k^al_a=-1, \;\; m^a\overline{m}_a=1
\end{equation}
and all other inner products vanish. To put it in other words, at
each point one takes a timelike plane, two vectors $k^a$ and $l^a$
lying along its real null directions, and two vectors $m^a$ and
$\mc^a$ lying along the complex conjugate null directions of the
orthogonal spacelike plane, these pairs of vectors satisfying the normalization conditions (\ref{normal conds}). We use the labels $\ah$, $\bh$ etc.\ for the tetrad indices. The basic variables of the
formalism are the spin coefficients ($\Gamma_{\ah\bh\ch}\equiv \ee_\ah{}^a\nabla_c(\ee_\bh)_a\,\ee_\ch{}^c=-\Gamma_{\bh\ah\ch}$)
\begin{eqnarray}
\kappa=\Gamma_{414},\quad \tau = \Gamma_{413},\quad \sigma= \Gamma_{411},\quad \rho=\Gamma_{412},\label{eq:Scalars1}\\
\nu = \Gamma_{233},\quad \pi = \Gamma_{234},\quad \lambda = \Gamma_{232},\quad \mu = \Gamma_{231},\label{eq:Scalars2}
\end{eqnarray}
the 9 independent components of the traceless part of the Ricci tensor $S_{ab}=R_{ab}-\frac{1}{4}Rg_{ab}$,
\begin{eqnarray}
&&\Phi_{00} = \frac{1}{2}S_{ab}k^ak^b, \quad \Phi_{22} =
\frac{1}{2}S_{ab}l^al^b,\label{Phi00}\\
&&\Phi_{01} = \frac{1}{2}S_{ab}k^am^b, \quad \Phi_{12} = \frac{1}{2}S_{ab}l^am^b,\\
&&\Phi_{02} = \frac{1}{2}S_{ab}m^am^b, \quad \Phi_{11} = \frac{1}{2}S_{ab}\left(k^al^b+m^a\overline{m}^b\right),\;\;\;\;\;\;\;\;
\end{eqnarray}
with $\Phi_{ji}=\overline{\Phi_{ij}}$, the multiple
\begin{equation}
\Pi\equiv \frac{R}{24}
\end{equation}
of the Ricci scalar,
and the 10 independent
components of the Weyl tensor $C_{abcd}$,
\begin{eqnarray}
&&\Psi_0  =  C_{abcd}k^am^bk^cm^d,\quad \Psi_4  =  C_{abcd}l^a\overline{m}^bl^c\overline{m}^d,\;\;\;\;\;\;\;\;\\
&&\Psi_1  =  C_{abcd}k^al^bk^cm^d,\quad \Psi_3  =  C_{abcd}l^ak^bl^c\overline{m}^d,\\
&&\Psi_2  =  C_{abcd}k^am^b\overline{m}^c l^d.\label{Psi2}
\end{eqnarray}

Changes of the tetrad leaving the null directions
spanned by $k^a$, $l^a$, $m^a$ and $\mc^a$ invariant, and at the
same time preserving the normalization conditions (\ref{normal
conds}), consist of boosts
\begin{equation}
\label{eq:boost}
k^a \rightarrow Ak^a, \; l^a\rightarrow A^{-1}l^a
\end{equation}
and spatial rotations
\begin{equation}
\label{eq:Rot}
m^a\rightarrow e^{i\theta}m^a.
\end{equation}
Quantities transforming under (\ref{eq:boost})-(\ref{eq:Rot}) as
\begin{equation}
\eta\rightarrow
A^{\frac{\we_p+\we_q}{2}}e^{i\frac{\we_p-\we_q}{2}\theta}\eta\label{eq:weightdef}
\end{equation}
are called {\em well-weighted of type} $\left(\we_p,\we_q\right)$ or  $(\we_p,\we_q)${\em-weighted} ({\em zero-weighted} in the case of type $(0,0)$).
They have boost-weight $\we_B(\eta)=\frac{\we_p+\we_q}{2}$ and
spin-weight $\we_S(\eta)=\frac{\we_p-\we_q}{2}$. One can check that
the GHP basic variables are well-weighted, their weights following from the definitions (\ref{eq:Scalars1})-(\ref{Psi2}) and (\ref{eq:boost})-(\ref{eq:weightdef}) - see also equation (7.36) in \cite{SKMHH}. E.g.\ $\we_B(\nu)=-2$, $\we_S(\nu)=-1$, implying $\nu$ is of type (-3,-1).
The following derivative operators are defined such that a well-weighted
quantity $\eta$ is transformed in a well-weighted quantity:
\begin{equation}
\label{eq:GHPderivs} D_\ah\eta=\ee_\ah(\eta)
+\Gamma_{34\ah}\we_B(\eta)\,\eta+\Gamma_{12\ah}\we_S(\eta)\,\eta.
\end{equation}
When $\eta$ is of type $(\we_p,\we_q)$ one can check that
\[
\we_B(D_\ah\eta)=\we_B(\eta)+\tilde{w}_B(\ah),\quad \we_S(D_A\eta)=\we_S(\eta)+\tilde{w}_S(\ah),
\]
where
\begin{eqnarray*}
\tilde{w}_B(\ah) =  \left\{\begin{array}{ll}
1, & \ah=4,\\
-1, & \ah=3,\\
0, & \ah=1,2
\end{array}\right.,\quad
\tilde{w}_S(\ah) =  \left\{\begin{array}{ll}
1, & \ah=1\\
-1, & \ah=2\\
0, & \ah=3,4.
\end{array}\right.
\end{eqnarray*}
One uses the notation
\begin{equation}
\eth\equiv D_1, \quad\eth'\equiv D_2, \quad \tho'\equiv D_3, \quad\tho\equiv D_4.
\end{equation}
Notice that the differential of zero-weighted scalars $f$ can be
expressed as
\begin{eqnarray}
\di_a f &=&-\tho'f k_a-\tho f l_a+\eth' f m_a+\eth f
\mc_a\label{dfweighted}\\
&=&-\ll(f)k_a-\kk(f)l_a+\mmc(f)m_a+\mm(f)\mc_a\label{dfnull}\\
&=&-\u(f)u_a+\v(f)v_a+\mmc(f)m_a+\mm(f)\mc_a,\;\;\;\;\;\;\;\;\;\label{dftimelike}
\end{eqnarray}
where $u^a$ and $v^a$ are related to $k^a$ and $l^a$ according to
(\ref{udef}), resp. (\ref{vdef}).

The basic (or `structure') equations of the GHP formalism are\\
(a) the commutator relations of the weighted derivatives, in the joint $D_\ah$ notation given by
\begin{eqnarray*}
\left[D_\ah,D_\bh\right]\eta = 2\Gamma^\ch_{[\ah\bh]}D_\ch\eta
+\we_B(\eta)\left(R_{34\ah\bh}+2\Gamma_{3\ch[\ah}\Gamma^\ch_{|4|\bh]}\right)\eta\\
+\we_S(\eta)\left(R_{12\ah\bh}+2\Gamma_{1\ch[\ah}\Gamma^\ch_{|2|\bh]}\right)\eta
+\tilde{w}_B(\bh)\Gamma_{34\ah}D_\bh\eta\\
+\tilde{w}_S(\bh)\Gamma_{12\ah}D_\bh\eta
-\tilde{w}_B(\ah)\Gamma_{34\bh}D_\ah\eta
-\tilde{w}_S(\ah)\Gamma_{12\bh}D_\ah\eta;
\end{eqnarray*}
(b) 12 complex Ricci identities (or `equations'), namely
\begin{equation*}
\ee_\ch(\Gamma_{\ah\bh\ddh})-\ee_\ddh(\Gamma_{\ah\bh\ch})=R_{\ah\bh\ch\ddh}-2\Gamma_{\ah\eh[\ch|}\Gamma^\eh_{\bh|\ddh]}-2\Gamma_{\ah\bh\eh}\Gamma^\eh_{[\ch\ddh]}
\end{equation*}
with $[\ah\bh]=[14],\,[23]$ (the complex conjugates corresponding to $[\ah\bh]=[24],\,[13]$);\\
(c) the Bianchi identities (or `equations')
\begin{equation*}
\ee_{[\fh}(R_{|\ah\bh|\ch\ddh]}) = -2 R_{\ah\bh\eh[\ch}\Gamma^\eh_{\ddh\fh]}+\Gamma^\eh_{\ah[\ch}R_{\ddh\fh]\eh\bh}-\Gamma^\eh_{\bh[\ch}R_{\ddh\fh]\eh\ah}.
\end{equation*}
One can show that, after writing the directional derivatives $\ee_\ah$ in terms of the weighted derivatives $D_\ah$, these basic equations (a)-(c) form a consistent, closed system of PDEs in the variables (\ref{eq:Scalars1})-(\ref{Psi2}) and with formal derivative operators $D_\ah$. Compared to the NP formalism, the 6 complex Ricci identities
which concern directional derivatives of the non-well-weighted NP spin coefficients $\alpha$, $\beta$, $\gamma$ and $\epsilon$ (corresponding to $[\ah\bh]=[12],\,[34]$) have been absorbed in the commutator relations (1). Explicitly, for a $(\we_p,\we_q)$-weighted scalar one gets
\begin{eqnarray}
&&[\tho,\tho'](\eta)=(\pi+\tauc)\eth(\eta)+(\pic+\tau)\eth'(\eta)\nonumber\\
&&\quad+(\k\nu-\pi\tau+\Pi-\Phi_{11}-\Psi_2)\we_p\,\eta\nonumber\\
&&\quad+(\kc\nuc-\pic\tauc+\Pi-\Phi_{11}-\Pc_2)\we_q\,\eta,\label{[tho,tho']}\\
&&[\eth,\eth'](\eta)=(\mu-\muc)\tho(\eta)+(\rho-\rhoc)\tho'(\eta)\nonumber\\
&&\quad+(\l\s-\mu\rho-\Pi-\Phi_{11}+\Psi_2)\we_p\,\eta\nonumber\\
&&\quad-(\overline{\l\s}-\muc\rhoc-\Pi-\Phi_{11}+\Pc_2)\we_q\,\eta,\label{[eth,eth']}\\
&&[\tho,\eth](\eta)=\pic\,\tho(\eta)-\k\tho'(\eta)+\rhoc\,\eth(\eta)+\sigma\eth'(\eta)\nonumber\\
&&\quad+(\k\mu-\s\pi-\Psi_1)\we_p\,\eta\nonumber\\
&&\quad+(\overline{\k\l}-\pic\rhoc-\Phi_{01})\we_q\,\eta,\label{[tho,eth]}
\end{eqnarray}
together with the equations obtained by applying the complex conjugate and/or prime dual operation to (\ref{[tho,eth]}). This {\em prime dual operation} is generated by interchanging $k^a\leftrightarrow l^a$ and $m^a\leftrightarrow \mc^a$, which comes down to
\begin{eqnarray}
&&\k\leftrightarrow -\nu,\quad \tau\leftrightarrow -\pi,\quad \s\leftrightarrow -\l,\quad \rho\leftrightarrow -\mu,\;\;\;\;\label{prime spin}\\
&&\Phi_{ij}\leftrightarrow \Phi_{2-i\,2-j},\quad \Psi_i\leftrightarrow \Psi_{4-i},\\
&&\tho\leftrightarrow\tho',\quad \eth\leftrightarrow\eth'.\label{prime deriv}
\end{eqnarray}
The interchange (\ref{prime deriv}) means that  $(\tho\eta)'=\tho'\eta'$ etc., and is due to (\ref{eq:GHPderivs}) and
\begin{eqnarray*}
&&\we_B(\eta')=-\we_B(\eta),\quad\we_S(\eta')=-\we_S(\eta),\\
&&\mbox{i.e.}\;\;\;\we_p(\eta')=-\we_p(\eta),\quad\we_q(\eta')=-\we_q(\eta).
\end{eqnarray*}
Regarding complex conjugation one has $\overline{\tho\eta}=\tho\bar{\eta}$, $\overline{\eth\eta}=\eth'\bar{\eta}$ and
\begin{eqnarray*}
&&\we_B(\bar{\eta})=\we_B(\eta),\,\we_S(\bar{\eta})=-\we_S(\eta),\\
&&\mbox{i.e.}\;\;\;\we_p(\bar{\eta})=\we_q(\eta),\quad\we_q(\bar{\eta})=\we_p(\eta).
\end{eqnarray*}
Explicitly, the 12 complex Ricci identities read
\begin{eqnarray}
&&\tho\tau-\tho'\k=(\tau+\pic)\rho+(\tauc+\pi)\s+\Phi_{01}+\Psi_1,\;\;\;\;\;\;\;\label{ghp1}\\
&&\eth\rho-\eth'\s=(\rho-\rhoc)\tau+(\mu-\muc)\k+\Phi_{01}-\Psi_1,\label{ghp2}\\
&&\tho\s-\eth\k=(\rho+\rhoc)\s+(\pic-\tau)\k+\Psi_0,\label{ghp3}\\
&&\tho\rho-\eth'\k=\rho^2+\s\sc-\kc\tau+\k\pi+\Phi_{00},\label{ghp4}\\
&&\tho'\s-\eth\tau=-\s\mu-\lc\rho-\tau^2+\k\nuc-\Phi_{02},\label{ghp5}\\
&&\tho'\rho-\eth'\tau=-\muc\rho-\l\s-\tau\tauc+\k\nu-2\Pi-\Psi_2\label{ghp6}
\end{eqnarray}
and their prime duals (\ref{ghp1})'-(\ref{ghp6})'.
Finally, the
Bianchi identities involve weighted derivatives of the Riemann tensor components. In full generality they are given in ref.\ \cite{SKMHH}, (7.32a-k), or \cite{Penrose1}, (4.12.36-41).

The formalism is especially suited for situations where two null
directions are singled out
by the geometry,
such that $k^a$ and $l^a$ can
be chosen along them. In particular, the Weyl tensor of a  Petrov type
$D$ spacetime  has precisely two PNDs;
choosing $k^a$ and $l^a$ along them is equivalent to condition
(\ref{Psis}), and a complex null tetrad
realizing this condition is called a {\em Weyl principal null
tetrad} (WPNT). When  (\ref{Phi_ij 0}) and (\ref{Psis}) are both satisfied, the Bianchi identities reduce to
\begin{eqnarray}
&&0=\s(2\Phi_{11}+3\Psi_2)-\lc\Phi_{00},\label{bi1}\\
&&\tho\Psi_2+\tho'\Phi_{00}+2\tho\Pi=\rho(2\Phi_{11}+3\Psi_2)-\muc\Phi_{00},\\
&&\tho\Phi_{11}+\tho'\Phi_{00}+3\tho\Pi=2(\rho+\rhoc)\Phi_{11}-(\mu+\muc)\Phi_{00},\;\;\;\;\;\;\;\;\\
&&\eth\Psi_2+2\eth\Pi=-\tau(2\Phi_{11}-3\Psi_2)+\nuc\Phi_{00},\\
&&\eth\Phi_{11}-3\eth\Pi=2(\tau-\pic)\Phi{11}-\nuc\Phi_{00}+\k\Phi_{22},\\
&&\eth\Phi_{00}=\k(2\Phi_{11}-3\Psi_2)-\pic\Phi_{00}\label{bi6}
\end{eqnarray}
and their prime duals (\ref{bi1})'-(\ref{bi6})'.

In general, the GHP formalism may be used to find a class of solutions, defined by a particular set of properties. One first translates these properties in terms of GHP variables, yielding (algebraic or differential) constraints on the system of basic equations, then recloses the resulting extended system (integrability analysis) and finally describes the corresponding metrics in terms of coordinates (integration). These coordinates are four suitable, functionally independent zero-weighted scalars $f$; they may be combinations of (derivatives of) basic variables, appearing in the reclosed system ${\cal S}$ itself, or `external' coordinates associated to HO vector fields due to Frobenius' theorem. The geometric duals of the null tetrad vectors, and hence the metrics $g_{ab}=-2k_{(a}l_{b)}+2m_{(a}\mc_{b)}$, are obtained by inverting (\ref{dfweighted}) for the chosen  $f$'s. Eventually the remaining equations of ${\cal S}$ are written in terms of these coordinates and the resulting PDE's are solved as far as possible. We refer to \cite{Edgar1} for enlightening discussions, and to e.g.\ \cite{EL} or this work for illustrations. In particular for Petrov type $D$ spacetimes, notice that zero-weighted combinations of {\em WPNT} spin coefficients and their weighted derivatives (e.g.\ $\mu\rho$ or $\eth'\tau$) are {\em scalar (Lorentz) invariants} $x$, which are thus annihilated by any present KVF $K^a$, ${\mathbf K}(x)={\cal L}_{\bf K} x=0$. This facilitates the detection of KVFs.
More generally, zero-weighted tensor fields $T_{ab\ldots}$, algebraically constructed
from the Riemann tensor, WPNT vectors and covariant derivatives thereof, are {\em invariantly-defined} by the geometry, and ${\cal L}_{\bf K}T_{ab\ldots}=0$.

\section{(Rigid) shearfree normality and
staticity of Petrov type $D$ spacetimes}

Consider (an open region of) a spacetime and a unit timelike vector field $u^a$ defined on it. Choose a null vector field $k^a$. At each point, $k^a$ and $u^a$ span a timelike plane $\Sigma$, the first null direction of which is spanned by $k^a$. Construct the null vector field $l^a$ by taking at each point the unique vector lying along the second null direction and satisfying $k^al_a=-1$. Then $u^a$ is decomposed as in (\ref{udef}), where $q=A^2$, $A=-(\sqrt{2}k^au_a)^{-1}$. The field $v^a$ defined in (\ref{vdef}) determines at each point the up to reflection unique unit spacelike vector lying in $\Sigma$ and orthogonal to $u^a$. The electric and magnetic parts of the Weyl tensor wrt $u^a$ can be decomposed as
\begin{eqnarray}
&&E_{ab}\equiv C_{acbd}u^cu^d
=(\Psi_2+\Pc_2)[v_av_b-\mc_{(a}m_{b)}]\nonumber\\
&&+\left[\frac{\Psi_4+q^2\Pc_0}{2q}m_am_b+2\frac{\Psi_3-q\Pc_1}{\sqrt{2q}}m_{(a}v_{b)}\right]\,+\mbox{c.c},\;\;\;\;\;\;\;\label{Edef}\\
&&H_{ab}\equiv\frac{\eta_{acmn}}{2}C^{mn}{}_{bd}u^cu^d
=\I(\Psi_2-\Pc_2)[v_av_b-\mc_{(a}m_{b)}]\nonumber\\
&&+\I\left[\frac{\Psi_4-q^2\Pc_0}{2q}m_am_b+2\frac{\Psi_3+q\Pc_1}{\sqrt{2q}}m_{(a}v_{b)}\right]\,+\mbox{c.c}.\;\;\;\;\;\;\;\label{Hdef}
\end{eqnarray}
If $u^a$ exists such that $H_{ab}=0$, the Weyl tensor is {\em purely electric} (PE) {\em wrt} $u^a$, the spacetime itself being also called PE. A criterion in terms of Weyl tensor concomitants, deciding whether this is the case, follows from the flow diagram 9.1 in \cite{SKMHH} and theorem 1 in \cite{McIntosh1}.

Suppose now that the spacetime admits a unit timelike vector field $u^a$ satisfying (\ref{ua;b}) - 
corresponding to an US, i.e. forming the tangent field of a shearfree and
vorticity-free cloud of test particles.
Within the GHP formalism based on $k^a$ and $l^a$ as introduced above, this is the case if and only if a (-2,-2)-weighted field $q$ exists satisfying
(\ref{lambda/sigma pi/tau})-(\ref{deriv q}). By virtue of these relations,
the $[\eth,\eth'](q)$ commutator relation yields (\ref{Psi2 real}), adding $2q[\overline{(\ref{ghp2})}-(\ref{ghp2})']$ to the $[\tho'-q\tho,\eth'](q)$ commutator relation gives $\Psi_3+q\Pc_1=0$ and the combination $q^2\overline{(\ref{ghp3})}-(\ref{ghp3})'+q[(\ref{ghp5})'-\overline{(\ref{ghp5})}]$ produces $\Psi_4-q^2\Pc_0=0$. Hence $H_{ab}=0$ from (\ref{Hdef}), and if we choose $k^a$ to be a (multiple) PND, $\Psi_0=0$ ($\Psi_0=\Psi_1=0$), then also $l^a$ is a (multiple) PND, $\Psi_4=0$ ($\Psi_4=\Psi_3=0$).
Hence the spacetime must be either conformally flat (all $\Psi_i$ zero), and then USs are always admitted (see e.g.\ (6.15)) in \cite{SKMHH}), or purely electric (PE) and of Petrov type $D$ or $I$, the Weyl tensor being PE wrt $u^a$.
For Petrov type $I$, there are 4 distinct PNDs, and $u^a$ is the up to reflection unique timelike vector lying along the intersection of the planes spanned by two particular pairs of PNDs. For Petrov type $D$, $k^a$ and $l^a$ can be taken to be the multiple PNDs, and $u^a$ lies in the plane $\Sigma$ spanned by them.

Propositions 4 in \cite{FerrSaezP1} and 16 in \cite{FerrSaezP2} imply intrinsic, easily testable criteria for deciding when a Petrov type I spacetime admits an US, resp.\ is static. Here we present likewise criteria in the Petrov type $D$ case. These criteria are invariant statements, in terms of GHP basic variables and weighted derivatives associated to an arbitrary WPNT. Given a Petrov type $D$ spacetime in coordinates, the determination of the PNDs, and hence the WPNTs, is straightforward and can be performed covariantly. It then suffices to fix one WPNT and calculate the appearing spin-boost covariant expressions by using definitions (\ref{eq:Scalars1})-(\ref{Psi2}) and (\ref{eq:GHPderivs}).
For complex $(2k,2k)$-weighted scalars ($k\in\mathbb{Z}$) $z=\Re(z)+i\,\Im(z)$ we mean with
$z>0$ ($z<0$) that $z$ is real and strict positive (negative) in the sequel.

It turns out that, given (\ref{Psis})-(\ref{Psi2 real}), the integrability conditions of (\ref{deriv q}) are identically satisfied. Thus we find: 
\begin{Proposition}\label{Prop: US} A Petrov type $D$ spacetime admits an US if and only if, wrt an arbitrary
WPNT, $\Psi_2$ is real and one of the following sets of conditions holds:
\begin{enumerate}
\item[1.] $\sigma\neq 0$, the scalar invariant $\lambda\sigma>0$, and
$q_0\equiv\l/\sic$ satisfies (\ref{lambda/sigma pi/tau})-(\ref{deriv q});
\item[2.]  $\rho\neq\rhoc$, the real scalar invariant $(\mu-\muc)(\rho-\rhoc)>0$, and
$q_0\equiv-(\mu-\muc)/(\rho-\rhoc)$ satisfies  (\ref{lambda/sigma pi/tau})-(\ref{deriv q});
\item[3.] $\lambda=\sigma=\mu-\muc=\rho-\rhoc=0$, the scalar invariant
$\kappa\nu\neq 0$ and one of the following situations occurs,
where $q_0$ defined in each subcase satisfies (\ref{deriv
q}) and where $b\equiv (\pi+\tauc)/\kappa$, $c\equiv
\nu/\kappa$: 
\begin{enumerate}
\item[3a.] $\Im(b)\Im(c)>0$ and $q_0\equiv\Im(c)/\Im(b)$ also
satisfies $q_0^2-\Re(b)q_0+\Re(c)=0$;
\item[3b.] $b=\Re(b)$, $c<0$ and $q_0\equiv(b+\sqrt{b^2-4c})/2$;
\item[3c.] $b>0$, $c>0$, $b^2\geq
4c$, and $q_0\equiv(b+\sqrt{b^2-4c})/2$ or $q_0\equiv
(b-\sqrt{b^2-4c})/2$;
\end{enumerate}
\item[4.] $\lambda=\sigma=\mu-\muc=\rho-\rhoc=0$, and either
\begin{enumerate}
\item[4.1]$\kappa=0\neq \nu$,
$(\pic+\tau)\nu>0$, and $q_0=\nu/(\pi+\tauc)$
satisfies (\ref{deriv q}), or
\item[4.2]$\kappa\neq0=\nu$,
$(\pi+\tauc)\kappa>0$ and $q_0=(\pi+\tauc)/\kc$
satisfies (\ref{deriv q});
\end{enumerate}
\item[5.] the WPNT directions are HO, i.e., (\ref{geod shearfree})-(\ref{pitau}) holds.
\end{enumerate}
\end{Proposition}

\noindent The subdivision of case 3 stems from a straightforward analysis of
the second equation of (\ref{lambda/sigma pi/tau}).
In cases 1, 2, 3a, 3b and 4 there is a unique US, whereas there may be one or two
USs in case 3c. Due to the number and nature of the equations (\ref{deriv q}) there is a one-degree freedom of USs in case 5, where the condition $\Psi_2=\bar{\Psi_2}$ can be dropped since it is implied by
the imaginary part of (\ref{ghp6}) + (\ref{ghp6})' and (\ref{geod shearfree})-(\ref{pitau}).
Important examples of
spacetimes satisfying criterion 5 are the Petrov type $D$ purely electric
Einstein spaces
and their `electrovac' generalizations (see
\cite{CzaporMcLen,DebeverMcLen} and $\S$~\ref{subsection:Einstein
spaces}) and all spacetimes with (pseudo-)spherical or planar
symmetry (which constitute the LRS
 II Lorentzian spaces, see \cite{StewartEllis}). These examples
all satisfy (\ref{Phi_ij 0}) on top of (\ref{geod
shearfree})-(\ref{pitau}) and are further characterized by
$\Phi_{00}=\Phi_{22}=(\Phi_{11}=)\,0$, resp.\ $\pi=\tau=\eth R=0$
(cf.\ \cite{Goode}).\\

The spacetime will admit a unit timelike vector field $u^a$
satisfying
\begin{equation}\label{ua;b non-expand}
u_{a;b}=-\dot{u}_au_b,
\end{equation}
corresponding to a rigid US or modeling a
rigid non-rotating cloud of test particles, if and only if a (-2,-2)-weighted field $q$ exists satisfying
(\ref{lambda/sigma pi/tau})-(\ref{deriv q}) and (\ref{mu=qrhoc}). Notice
that, given (\ref{mu=qrhoc}), the third equation of (\ref{lambda/sigma pi/tau}) is identically satisfied. Hence we have
\begin{Proposition}\label{Prop: rigid US}
A Petrov type $D$ spacetime admits a rigid US if and only if, wrt an arbitrary WPNT,
$\Psi_2$ is real and one of the following sets of conditions holds:
\begin{enumerate}
\item[1'.] condition 1 with the third equation of (\ref{lambda/sigma pi/tau}) replaced by (\ref{mu=qrhoc});
\item[2'.] the scalar invariant ${\mu\rho>0}$ and
$q_0\equiv\mu/\rhoc$ satisfies (\ref{lambda/sigma pi/tau})-(\ref{deriv q});
\item[3'-5'.] conditions 3-5 with $\mu-\muc=\rho-\rhoc=0$
replaced by $\mu=\rho=0$.
\end{enumerate}
In case 5', the spacetime possesses geodesic, shearfree and non-diverging PNDs ($\k=\s=\rho=0$, $\nu=\l=\mu=0$) - thus belonging to Kundt's class - and HO Weyl principal
complex null directions ($\l=\s=\pi+\tauc=0$), and admits a one-degree freedom of rigid USs.
\end{Proposition}

The spacetime is static if and only if it admits a HO timelike
KVF. An equivalent characterization was given by
Ehlers and Kundt~\cite{EhlersKundt}: the spacetime is static if and
only if a unit timelike vector field $u^a$ exists for which shear,
vorticity and expansion scalar vanish, i.e.  (\ref{ua;b non-expand})
holds, and for which the acceleration $\dot{u}^a$ is
Fermi-propagated along the integral curves of $u^a$:
\begin{equation}\label{static cond}
\ddot{u}_{[a}u_{b]}=0.
\end{equation}
The field $u^a$ is then parallel to a (HO and timelike) KVF and identified with a congruence of static observers. By a long but straightforward calculation, thereby
simplifying expressions by means of (\ref{lambda/sigma
pi/tau})-(\ref{deriv q}), (\ref{mu=qrhoc}), (\ref{ghp1}), (\ref{ghp1})' and the
$[\tho,\tho'](q)$ commutator relation, one shows that the extra
condition (\ref{static cond}) is equivalent to
\begin{eqnarray}
&&(q\k+q^{-1}\nuc)(\tho q+\sqrt{2q})-2\tho\nuc+2q\tho\tau\nonumber\\
&&\qquad
+\Phi_{12}-q\Phi_{01}=0,\label{cond1}\\
&&\tho\tho
q=\pi\tau+\pic\tauc-q(\k\pi+\kc\pic)-q^{-1}(\nu\pic+\nuc\pi)\nonumber\\
&&\qquad+2\Phi_{11}-\frac{R}{12}+2\Psi_2.\label{cond2}
\end{eqnarray}
In case 5' above, the Ricci equations (\ref{ghp1}), (\ref{ghp4}) and (\ref{ghp4})'  yield
$\tho\tau=\Phi_{01}$ and $\Phi_{00}=\Phi_{22}=0$, and so
(\ref{cond1})-(\ref{cond2}) reduces to
\begin{eqnarray}
&&\Phi_{12}+q\Phi_{01}=0,\label{Phi12/01}\\
&&\tho\tho
q=-2\tau\tauc+2\Phi_{11}-\frac{R}{12}+2\Psi_2.\label{thothoq}
\end{eqnarray}
In the subcase $\Phi_{01}=\Phi_{12}=0$ of (\ref{Phi12/01}), the $[\tho,\tho']$,
$[\tho,\eth]$ and $[\tho,\eth']$ commutators applied to $q$ yield
\begin{eqnarray}
&& \tho'\tho q =-q\tho\tho q+(\tho q)^2,\label{tho'thoq}\\
&&\eth\tho q=\tau \tho q,\quad \eth'\tho q=\tauc \tho
q.\label{eththoq}
\end{eqnarray}
The compatibility requirement of (\ref{thothoq})-(\ref{eththoq})
with the commutator relations for $\tho q$ gives the single
condition
\begin{eqnarray}\label{thoR}
\tho' R+q\tho R=0.
\end{eqnarray}
According to the Sach's star dual~\cite{GHP} of the LRS criterion in
\cite{Goode}, the subcase $\tho'R=\tho R=0$ of (\ref{thoR}) precisely
corresponds to a boost-isotropic spacetime with $\pi+\tauc=0$. From
the above we conclude:
\vspace{.3cm}
\begin{Proposition}\label{Prop: static}
A Petrov type D spacetime is static if
and only if, wrt an arbitrary WPNT, one of the following sets of
conditions holds:
\begin{enumerate}
\item[1''-4''.] $\Psi_2$ is real, conditions 1'-4' hold and $q_0$ additionally satisfies
(\ref{cond1})-(\ref{cond2});
\item[5''a.] condition 5' holds, the scalar invariant ${\Phi_{01}\Phi_{21}<0}$
and $q_0\equiv-\Phi_{12}/\Phi_{01}$ satisfies (\ref{deriv q}) and (\ref{thothoq});
\item[5''b.] condition 5' holds, $\Phi_{01}=\Phi_{21}=0$, the scalar invariant ${(\tho'R)(\tho R)<0}$
and $q_0\equiv -\tho' R/\tho R$ satisfies (\ref{deriv q}) and (\ref{thothoq});
\item[6''.] the spacetime is (locally) boost-isotropic and $\pi+\tauc=0$.
\end{enumerate}
The HO timelike KVF directions are parametrized by two constants in case 6''~\footnote{The two constants result from the integrations of (\ref{thothoq})-(\ref{eththoq}), giving $\tho q$, and consecutively of (\ref{deriv q}).}, are 1 or 2 in number in case 3''c and are unique in all other cases.
\end{Proposition}

\end{document}